\documentclass[iop,apj,numberedappendix,twocolappendix]{emulateapj_cmcnally}
\usepackage{graphicx,amsmath,url}
\usepackage{times}
\usepackage{natbib}
\usepackage{bm}
\usepackage{hyperref}
\usepackage[normalem]{ulem}

\newcommand{\V}{{\bm v}}
\newcommand{\OMEGA}{\mathbf{\Omega}}
 
\usepackage{color}
\definecolor{brown}{rgb}{0.42,0.24,0.07}
\definecolor{darkgreen}{rgb}{0.0,0.6,0.00}
\definecolor{purple}{rgb}{0.7,0.0,0.7}

\begin{document}

\title{On Vertically Global, Horizontally Local Models for  Astrophysical Disks}
\shorttitle{On Vertically Global Models for Disks}
\author{Colin~P.~McNally and Martin~E.~Pessah}
\shortauthors{McNally \& Pessah}
\affiliation{Niels Bohr International Academy, Niels Bohr Institute, Blegdamsvej 17, DK-2100, Copenhagen~\O, Denmark; cmcnally@nbi.dk, mpessah@nbi.dk}

\begin{abstract}
Disks with a barotropic equilibrium structure, for which the pressure is only a function of the
density, rotate on cylinders in the presence of a gravitational
potential, so that the angular frequency of such a disk is independent of height.
Such disks with barotropic equilibria can be { approximately} modeled using
the shearing box framework, representing a small disk volume with height-independent 
angular frequency.
If the disk is in baroclinic equilibrium,
 the angular frequency does in general depend on height and
it is thus necessary to go beyond the standard shearing box approach.
 In this paper, we show that given a global disk model, it is 
possible to develop { approximate} models that are local in horizontal 
planes { without an expansion in height}
with shearing-periodic boundary conditions. 
We refer to the resulting framework as the vertically global
shearing box (VGSB).
These models can be non-axisymmetric for 
globally barotropic equilibria but should be axisymmetric for 
globally baroclinic equilibria.
We provide explicit equations for this VGSB
 which can be implemented in standard magnetohydrodynamic
codes by generalizing the shearing-periodic
boundary conditions to 
allow for a height-dependent angular frequency and shear rate.
{ We also discuss the limitations that result from
the radial approximations that are needed in order to impose height-dependent
shearing periodic boundary conditions.}
We illustrate the potential of this framework
by studying a vertical shear
instability
and examining the modes associated 
with the magnetorotational instability.
\end{abstract}


\section{Introduction}
\label{sec_intro}

Astrophysical disks play a crucial role in the formation, evolution,
and fate of a wide variety of celestial objects, by mediating the
transport of mass, energy, and angular momentum.  Building realistic
disk models is of fundamental importance for understanding, for
example, protoplanetary disks around young stars, accretion flows onto
stellar compact objects and active galactic nuclei, as well as the 
interstellar medium in galactic disks.  The large
dynamic range involved makes it particularly challenging to produce
detailed global numerical simulations of these systems. Moreover, while 
global models allow us to investigate large-scale phenomena, local models 
with a hierarchy of increasingly complex microphysics have proved critical 
to elucidating the processes that are crucial at small scales. Because of 
this, several types of local approximations have been employed for studying 
astrophysical disks.

\citet{hill1878} pioneered the use of a local approximation to study
the dynamics of particles orbiting a host system and subject to
encounters with a perturber, and used it to study the motion of the
Moon.  \citet{1953ApJ...118..106S} and \citet{1965MNRAS.130..125G}
applied these ideas to galactic disks using the concept
of a locally shearing coordinate system. This approach constitutes the
basis of the {shearing sheet} framework which has been widely used 
to study the dynamics of orbiting particles and planetesimals, as
well as local processes in hydrodynamic and magnetohydrodynamic disks.  The
implementation of the concept behind the shearing sheet, with an
appropriate shear-periodic radial-boundary condition
\citep{1988AJ.....95..925W}, forms a computational model used for
studying local disk dynamics referred to as the {shearing box}
\citep{1995ApJ...446..741B, 1995ApJ...440..742H}.

Shearing box models solve the equations of motion for the fluid in a
local cartesian frame co-rotating with the disk at a fiducial radius.
In the standard framework, the differential rotation of the disk is
locally accounted for with a height-independent angular frequency.
This is appropriate for disks with a barotropic equilibrium
for which the pressure is only a function of density and
thus rotate on cylinders. The shearing box framework 
 relies on a
first order expansion of the steady bulk flow in the radial direction,
which is the highest order compatible with shearing periodic boundary
conditions. Depending on whether zeroth or first order expansions are
considered for the gravitational field in the direction perpendicular
to the disk, usually denoted with the coordinate $z$, this leads to
the so called unstratified \citep{1995ApJ...440..742H,
1996ApJ...464..690H} or stratified
\citep{1995ApJ...446..741B,1996ApJ...463..656S} shearing box models.
{ This stratified, compressible shearing box is what we will refer to in this paper
 as the standard shearing box (SSB). }
These approximations are appropriate when the disk is thin and the
vertical scales of interest are small compared to the fiducial disk
radius. 
There have been works that retained the correct expression for
the vertical component of the gravitational field
\citep{1997ASPC..121..766M,1999ApJ...514L..99K}, allowing for larger
vertical domains to be considered. However, this generalization does
not allow to study of disks with baroclinic equilibria without relaxing the
assumption that the angular frequency is height-independent. 
In the
early formulation of the shearing sheet, \citet{1965MNRAS.130..125G}
avoided making any approximation in height ($z$), which is possible
when considering barotropic equilibrium structures.

Astrophysical disks with baroclinic equilibrium structure, for which the pressure is not 
solely a function of density, posses angular frequency profiles
that depend in general on height, especially if these are not thin.
Therefore, building a framework to study these disks demands going
beyond the SSB, where the assumption that the 
angular frequency is height-independent is rooted deep.
In this work, we generalize the SSB by considering
the full height-dependence of a steady state, axisymmetric bulk flow to 
leading order in radius, without making any expansion in the vertical 
coordinate. 
We show that given a global disk model, it is 
possible to develop 
{ approximate models}
that are local in horizontal 
planes and global in height and are amenable to shearing-periodic
boundary conditions. These models can be non-axisymmetric for 
 disks with a barotropic global equilibrium but should be axisymmetric for 
disks with a baroclinic global equilibrium.
We term the resulting framework the vertically global shearing 
box (VGSB).
{ The terminology of `vertically global' specifically and solely refers 
to the approach of never making a expansion in the vertical direction in the derivation. 
This yields a model which allows us to account for the vertical variation of gravity
without approximations and the possible presence of vertical shear. As we discuss below, 
the radial expansions that are needed to apply height-dependent shearing periodic boundary 
conditions do limit the range of vertical scales which can be modeled approximately.
In spite of its limitations, the VGSB formalism provides a novel framework 
that goes beyond the SSB and promises to provide a bridge
between strictly local and fully global approaches to model astrophysical disks. 

}

The paper is organized as follows.
We derive the equations involved in the VGSB framework 
in Section~\ref{sec_derivation}, providing some of the 
algebraic details in Appendix \ref{sec_velocitysplitting}.
We state the final form of the  VGSB equations and discuss its novel features 
in Section~\ref{sec_final_eqs}. For convenience, we provide a self-contained summary 
of the  VGSB equations that can be incorporated in magnetohydrodynamic codes
in Appendix \ref{sec_vgsbsummary}.
We use this new framework to explore 
the behavior of two important instabilities in a baroclinic context.
We demonstrate that a linear vertical shear instability (VSI),
akin to those studied by \cite{1967ApJ...150..571G,1968ZA.....68..317F} 
and \cite{2013MNRAS.435.2610N}, appears in the VGSB in Section~\ref{sec_vsi}.
We examine some basic aspects of the magnetorotational instability (MRI) in the VGSB 
in Section~\ref{sec_mri}.
{ We conclude by briefly discussing the limitations and 
several potential applications of the VGSB in Section~\ref{sec_discussion}.
In addition, in Appendix~\ref{sec:neglecting_curvature} we outline some details 
related to neglecting curvature terms, in Appendix~\ref{sec:nondimensional_hydro_momentum} 
we analyze the hydrodynamic momentum equation in order to assess under what conditions
it is acceptable to discard radial pressure gradients in the SSB and VGSB, and in 
Appendix~\ref{sec:potvort} we discuss issues related to potential vorticity in shearing boxes.}

\section{Equations of Motion}
\label{sec_eq_motion}

We are concerned with the equations of ideal magnetohydrodynamics, 
in cylindrical coordinates $(r,\phi,z)$, in a reference frame 
rotating with angular frequency $\OMEGA_{\rm F}=\Omega_{\rm F} \hat{\bm{z}}$, i.e.,
\begin{align}
&\frac{\partial \rho}{\partial t} +\nabla\cdot\left( \rho \bm{v} \right) = 0 \,, \label{eq_cont} \\
&\frac{\partial \bm{v}}{\partial t} + \left(\bm{v}\cdot\nabla\right) \bm{v} =  
\Omega_{\rm F}^2 r\bm{\hat{r}} -2\bm{\Omega}_{\rm F}\times\bm{v}  \nonumber\\
  &\qquad  
    -\nabla\Phi - \frac{\nabla P}{\rho} +\frac{1}{\rho} \bm{J}\times \bm{B} \,, \label{eq_mom} \\
&
\frac{\partial \bm{B}}{\partial t} =  \nabla \times ({\bm v} \times {\bm B}) \,, 
     \label{eq_induc}\\
& \frac{\partial e}{\partial t} + \nabla\cdot\left( e \bm{v} \right)  = - P \left(\nabla \cdot \bm{v}\right)\, . \label{eq_energy}
\end{align}
Here,
$\rho$ is the mass density, 
$\bm{v}$ is the fluid velocity in the rotating frame, 
$\bm{B}$ is the magnetic field, with $\nabla \cdot {\bm B}=0$,
$e$ is the internal energy density,
$P(\rho,e)$ is the pressure determined through an equation of state, and 
$\Phi(r,z)$ is the gravitational potential, which is assumed to be cylindrically 
symmetric, but not necessarily spherical.
The current density is $\bm{J}\equiv \nabla\times \bm{B}/ \mu_0$,  
with $\mu_0$ a constant dependent on the unit system adopted.

Fluid flows described by these equations are subject to
conservation laws. It is thus important to understand under what 
circumstances these properties are satisfied by the equations 
describing the local dynamics involving expansions of the
original set of equations. It is easy to show that the approximations 
embodied in the standard isothermal shearing box are such that
the vortex lines of an inviscid flow are frozen into the fluid 
(Kelvin's Circulation Theorem) and that the magnetic flux is also 
frozen into the fluid in the absence of magnetic dissipation 
(Alfv\'{e}n's Frozen-in Theorem). 

Understanding under what conditions 
these properties also hold for the equations of motion
that result from invoking a local approximation
of global disk models which have a baroclinic 
equilibrium structure is more subtle. Here, 
we state the general versions of the aforementioned
theorems in order to prepare the ground to address these issues
in subsequent sections. 
These conservation theorems can be derived by calculating the 
Lagrangian derivative of the fluxes associated with the 
vorticity and the magnetic field. It is thus useful to recall that,
see e.g., \citep{2013EJPh...34..489B},
any vector field $\bm Q$ satisfies 
\begin{align}
\label{eq:flux_advection}
\Bigg(\frac{\partial}{\partial t}& + \bm{v} \cdot \nabla\Bigg) 
\int_{\rm S} \bm{Q} \cdot \bm{dS}  
\\ \nonumber
&=\int_{\rm S} 
\left[ \frac{\partial \bm{Q}}{\partial t}
- \nabla \times (\bm{v} \times \bm{Q}) 
+ \bm{v} (\nabla \cdot \bm{Q}) \right] \cdot \bm{dS} \,,
\end{align}
where the integral is carried out over any open surface 
${\rm S}$ advected by the flow with velocity ${\bm v}$. 

\subsection{Kelvin's Circulation Theorem}

The momentum equation (\ref{eq_mom}) for an 
an inviscid, unmagnetized, barotropic flow
in the rotating frame is given by
\begin{align}
(\partial_t + \bm{v}\cdot\nabla) \bm{v} = \Omega_{\rm F}^2  r\bm{\hat{r}} -2\bm{\Omega}_{\rm F}\times\bm{v} - \nabla\Theta  \,,
\label{eq_mom_barotropic}
\end{align}
where $\Theta = \Phi + h$ is the generalized gravito-thermal potential, 
where $h$ is the enthalpy, with $dh=dP/\rho$.
The equation governing the evolution of the vorticity is thus
\begin{align}
\partial_t (\nabla \times \bm{v}) =  \nabla \times [{\bm v} \times (\nabla \times {\bm v} + 2 {\bm \Omega}_{\rm F})]  \,,
\end{align}
which preserves the solenoidal character of the vorticity.
This implies that, by virtue of Equation~(\ref{eq:flux_advection})
with $\bm{Q} = \nabla \times \bm{v} + 2 {\bm \Omega}_{\rm F}$, 
vortex lines are frozen into the fluid, 
i.e., the flow preserves the circulation $\Gamma$
\begin{align}
(\partial_t + \bm{v} \cdot \nabla) \Gamma =  0 \,,
\label{eq:kelvin_theorem}
\end{align}
with
\begin{align}
\Gamma \equiv 
\int_{\rm S} (\nabla \times {\bm v} + 2 {\bm \Omega}_{\rm F})  \cdot \bm{dS} =
\oint_{\rm L} ({\bm v} + {\bm \Omega}_{\rm F} \times {\bm r}) \cdot d{\bm l}   \,,
\end{align}
where ${\rm L}$ is a closed contour, delimiting the open surface
${\rm S}$, advected by the flow with velocity ${\bm v}$.

\subsection{Alfv\'{e}n's Frozen-in Theorem}
 
The induction equation~(\ref{eq_induc}) 
preserves the divergence of the magnetic field, i.e., 
\begin{align}
(\partial_t + \bm{v} \cdot \nabla) ({\nabla\cdot \bm B})= 0 \,.
\end{align}
This implies that, provided that $\nabla\cdot \bm B =0$ at some initial time,
the magnetic flux remains frozen into the fluid, i.e., 
\begin{align}
(\partial_t + \bm{v} \cdot \nabla)   \Phi_{\rm B} = 0 \,,  
\label{eq:alfven_theorem}
\end{align}
with
\begin{align}
\Phi_{\rm B} = \int_{\rm S} \bm B \cdot \bm{dS}\, ,
\end{align}
where the integral is carried out over any open surface 
${\rm S}$ advected by the flow with velocity ${\bm v}$.
This follows from Equation~(\ref{eq:flux_advection})
with $\bm{Q} = \bm{B}$ and the induction equation
(\ref{eq_induc}).

\section{The Vertically Global, Horizontally Local Approximation}

\label{sec_derivation}

We seek to derive a set of equations that describe the local dynamics
of the magnetized fluid with respect to a known steady state bulk 
flow around a point co-rotating  with the disk at a distance $r_0$. 
Here, we outline the steps of the derivation, which is carried out 
in detail below.

\begin{enumerate}
\item We find a suitable steady flow and background equilibrium,
which enables the derivation of exact equations of motion for the 
departures from this solution.

\item We transform to a locally cartesian coordinate system. We expand 
the bulk flow and background equilibrium to leading order in the radial 
direction leaving the direction perpendicular to the disk midplane 
unaltered.

\item We determine under which circumstances the resulting equations
are amenable to being solved with shearing-periodic boundary conditions,
which could depend on height for disk models with baroclinic global equilibria.

\end{enumerate}

We show explicitly these steps for the momentum and induction 
equations, while we state the results for the continuity equation 
and energy equation that are simpler to work with.

\begin{figure*}
\includegraphics[width=\textwidth]{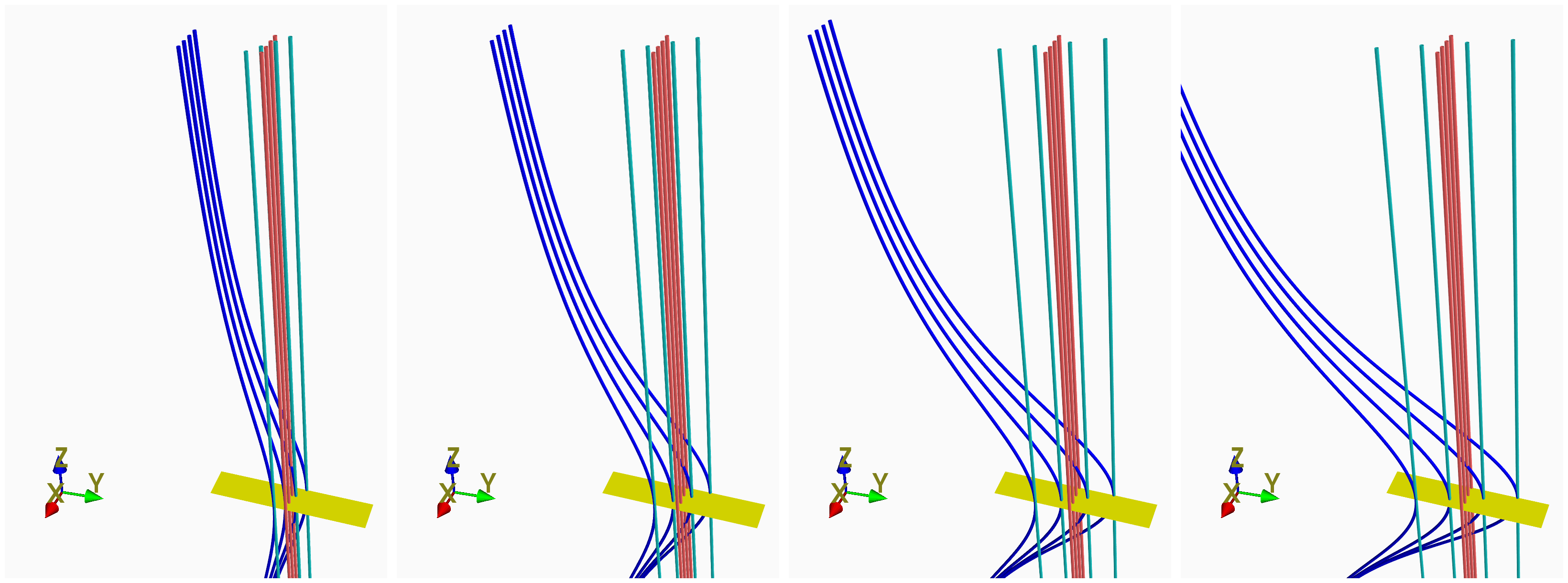}
\caption{
Rendering of Lagrangian tracer lines initially placed at $y=0$ in the steady state 
flow of the VGSB and standard shearing box. The VGSB provides a horizontally local, 
vertically global representation of a baroclinic equilibrium disk model with a 
cylindrical temperature dependence,
which is described in more detail in Section~\ref{sec_cylindrical_VGSB}, 
with a sound speed to rotation speed ratio of
$c_{{\rm s}0}/v_{{\rm K}0}=0.1$, $p=-1.5$, and $q=-1$.
The standard shearing box shown is for a Keplerian thin disk.
The panels corresponds to $t\Omega_0(0)= 0.5, 1.0, 1.5, 2.0$, from left to right.
The yellow surface marks the midplane  of the shearing box and the top of the lines is at $z=3r_0$.
{\em Red:} initial position of tracer lines. {\em Blue:} evolution of 
tracer lines in the VGSB following the flow $\bm{V}(x,z)=\left[V_0(z)+S_0(z)\,x\right]\bm{\hat{y}}$ 
from Equations~(\ref{eq_cylvgsbomega})--(\ref{eq_cylvgsbS}).
{\em Green:}  evolution of tracer lines in the standard shearing box following 
the flow $\bm{V}(x,z)=-(3/2) \Omega_0(0) \, x \, \bm{\hat{y}}$.  
}
 \label{fig_tubes1}
\end{figure*}

\subsection{Equations for Departures from Background Equilibrium}

\subsubsection{Bulk Flow and Background Equilibrium}

\label{sec_subtracting_bulk}

As in the SSB, we first seek a steady background flow.
In this case, we will not make any a priori assumptions about the $z$-dependence
of the angular frequency. 

We begin by noting that the momentum equation~(\ref{eq_mom}) 
admits a force-free (magnetic fields playing no role), 
steady state solution $\bm{V} \equiv V(r,z) \hat{\bm{\phi}}$, 
with
\begin{align}
V(r,z)  = r\left[\Omega(r,z) - \Omega_{\rm F}  \right] \,,
\label{eq_Vrz}
\end{align}
where the angular frequency is
\begin{align}
\Omega^2(r,z) \equiv \frac{1}{r \rho_h}\frac{\partial P(\rho_h,e_h)}{\partial r} 
  + \frac{1}{r}\frac{\partial \Phi}{\partial r}  \,, 
  \label{eq_omegarz}
\end{align}
and the vertical, hydrostatic pressure gradient satisfies
\begin{align}
\frac{1}{\rho_h}\frac{\partial P(\rho_h,e_h)}{\partial z} = - \frac{\partial \Phi}{\partial z} \,.
\label{eq_gradpz}
\end{align}
Here, $\rho_h(r,z)$ and $e_h(r,z)$ are the mass density and energy density associated
with the steady state background flow.

\subsubsection{Momentum and Induction Equations}

Using Equations~(\ref{eq_omegarz})~and~(\ref{eq_gradpz}) we can recast the 
gravitational force in the momentum equation~(\ref{eq_mom}) in terms of the
angular frequency and the pressure gradient, both corresponding to the steady
state bulk flow. We obtain
\begin{align}
\left( \frac{\partial }{\partial t} + \bm{v}\cdot\nabla \right) \bm{v} &=  \left[ \Omega_{\rm F}^2 -\Omega^2(r,z) \right]r \bm{\hat{r}} -2\bm{\Omega}_{\rm F}\times \bm{v} \nonumber    \\
 +\frac{\nabla P(\rho_h,e_h) }{\rho_h}  & - \frac{\nabla P(\rho,e)}{\rho} +\frac{1}{\rho} \bm{J}\times \bm{B} \,. \label{eq_mom_before_split}
\end{align}

The velocity field describing the departure from the bulk flow satisfying Equations~(\ref{eq_Vrz})--(\ref{eq_gradpz})
\begin{align}
\bm{w} \equiv \bm{v} - V(r,z)\hat{\bm{\phi}} \, , 
\end{align}
evolves according to the momentum equation,\footnote{
We provide the algebraic details of this derivation in Appendix~\ref{sec_velocitysplitting}.}
\begin{align} 
\frac{\partial \bm{w}}{\partial t} &  +\left[ \Omega(r,z) - \Omega_{\rm F}\right]  
\left(
\frac{\partial w_r}{\partial \phi} \hat{\bm{r}} +
\frac{\partial w_\phi}{\partial \phi} \hat{\bm{\phi}} +
\frac{\partial w_z}{\partial \phi} \hat{\bm{z}}
\right)
     + \bm{w}\cdot\nabla\bm{w}        
     \nonumber \\
      &+ w_r r \frac{\partial \Omega(r,z) }{\partial r} \hat{\bm{\phi}} + w_z \frac{\partial r \Omega(r,z)}{\partial z} \hat{\bm{\phi}} 
      +2\bm{\Omega}(r,z)\times \bm{w}  \nonumber \\
      & =   \frac{\nabla P(\rho_h,e_h)}{\rho_h} - \frac{\nabla P(\rho, e)}{\rho} +\frac{1}{\rho} \bm{J}\times \bm{B}  \, . \label{eq_mom_w}
\end{align}
This equation is exact and it displays the particular feature that the last term 
on the left-hand side resembles
the Coriolis acceleration, with one important difference. The angular frequency involved is not the
fixed angular frequency of the rotating frame, $\bm{\Omega}_{\rm F}$, but rather 
the angular frequency of the steady state flow, $\bm{\Omega}(r,z)$.
As a quick check, note that if
$P(\rho,e) = P(\rho_h,e_h)$ and $\bm{J}\times \bm{B}=0$
then there are no departures from the steady state bulk flow, i.e., $\bm{w}=0$.

Taking the induction equation~(\ref{eq_induc}) and replacing the velocity 
field with $\bm{v} \equiv V(r,z)\hat{\bm{\phi}} + \bm{w}$ yields, after 
some algebra,
\begin{align}
&\frac{\partial {\bm B} }{\partial t} 
+ \left[ \Omega(r,z) - \Omega_{\rm F}\right] \left(\frac{\partial B_r}{\partial \phi} \hat{\bm r} 
+ \frac{\partial B_\phi}{\partial \phi} \hat{\bm \phi} 
+ \frac{\partial B_z}{\partial \phi} \hat{\bm z}  \right)  \nonumber\\
&
= B_r r \frac{\partial \Omega(r,z)}{\partial r}\hat{\bm \phi}  
+ B_z\frac{\partial r\Omega(r,z)}{\partial z}\hat{\bm \phi}
+ \nabla \times ( {\bm w} \times {\bm B} ) \ . 
\label{eq_splitinduction}
\end{align}

\subsubsection{Background-flow Advection and Shear Rate}

The fact that the speed of the background flow $\bm{V} \equiv V(r,z) \hat{\bm{\phi}}$
depends, in general, on height implies that the departures from the bulk flow will 
be advected and sheared in a height-dependent way. This motivates the definition of the 
advection operator
\begin{align}
\mathcal{D} &\equiv \frac{\partial}{\partial t} + \left[\Omega(r,z) - \Omega_{\rm F}\right] \frac{\partial}{\partial \phi} \,, \label{mathcal_D}
\end{align}
which is defined so that it acts on scalar fields, 
such as the density $\rho$, and on each of the components 
of a vector field, e.g., $\bm{w}$ and $\bm{B}$, but not on 
the unit coordinate-vectors, i.e., $\mathcal{D}\{\hat{\bm{r}}, \hat{\bm{\phi}}\} = 0$. It is also convenient to define the shear rate
\begin{align}
S(r,z) &\equiv r \frac{\partial \Omega(r,z)}{\partial r} \,.
\end{align}

Using these definitions, Equations (\ref{eq_mom_w}) and (\ref{eq_splitinduction}) become, without approximations, 
\begin{align}
\left(\mathcal{D}+ \bm{w} \cdot \nabla\right) \bm{w} = &-2\Omega(r,z)\hat{\bm{z}}\times\bm{w} - S(r,z) w_r \bm{\hat{\phi}} \nonumber\\
- w_z\frac{\partial V(r,z)}{\partial z}&\bm{\hat{\phi}} +  \frac{\nabla P(\rho_h, e_h)}{\rho_h}   
- \frac{\nabla P(\rho, e)}{\rho}  + \frac{1}{\rho}\bm{J}\times \bm{B} 
\label{momentum_w} \,,
\end{align}
\begin{align}
\mathcal{D}{\bm B}=
&
  S(r,z) B_r \hat{\bm \phi}
+ B_z\frac{\partial V(r,z) }{\partial z}\hat{\bm \phi}
+ \nabla \times ( {\bm w} \times {\bm B} ) \,.
\label{eq:induction_D_S}
\end{align}

\subsection{Local Approximation in Horizontal Planes}
\label{sec_approx_in_horiz_planes}

We now seek to derive a set of equations of motion which is local 
in radius and azimuth by expanding Equation~(\ref{momentum_w}) 
and (\ref{eq:induction_D_S}) around a fiducial point $\bm{r}_0 = (r_0,\phi_0,0)$.
In order to simplify this task, we choose a reference 
frame that corotates with the bulk flow
at radius $r_0$, i.e., 
\begin{align}
\Omega_{\rm F} = \Omega(r_0,0) \, .
\end{align}
We also adopt a coordinate frame centered at $\bm{r}_0$ 
with locally cartesian coordinates $\bm x = (x,y,z)$, 
such that $x=r-r_0$ and $y \equiv r_0(\phi-\phi_0)$, with
$x/r_0\ll 1$ and $y/r_0\ll 1$. 
In this locally cartesian frame, the differential vector operators 
are well approximated by their cartesian versions, provided that 
the radial coordinate versor $\bm{\hat{r}}(\phi) \approx \bm{\hat{r}}(\phi_0)$
\footnote{This approximation, which is valid when the physical extent in horizontal planes is small, 
leads to neglecting terms related to the curvilinear character
of the cylindrical coordinate system originally chosen
to describe the bulk flow as $\bm{V} \equiv V(r,z) \hat{\bm{\phi}}$.
See Appendix \ref{sec:neglecting_curvature} for more details.} 
Because of the axisymmetric character of the background flow, in what follows,
we choose $\phi_0 =0$ without loss of generality.

In this locally cartesian coordinate system, we can expand to leading
order in $(x,y)$ the various functions appearing in the momentum and 
induction equations, Equation~(\ref{momentum_w}) and (\ref{eq:induction_D_S}), 
respectively. 

\subsubsection{Approximation of the Bulk Flow}

The local approximations of the angular frequency, the bulk flow in 
Equation~(\ref{eq_Vrz}), and the advection operator in 
Equation~(\ref{mathcal_D}) yield
\begin{align}
\Omega(x,z) &\equiv \Omega_0(z) +\left. \frac{\partial \Omega(r,z)}{\partial r}\right|_{r=r_0} x 
\,, \label{eq_omegarz_exp}  \, \\
V(x,z) & \equiv V_0(z) + S_0(z) \,x 
\,,
\label{eq_local_bulk_flow}
\, \\
\mathcal{D}_0 &\equiv \frac{\partial}{\partial t} +  \left[V_0(z) + S_0(z) x \right] \frac{\partial}{\partial y} \label{eq_def_D0} \ .
\end{align}
Here, we have defined the local, height-dependent angular frequency,
bulk flow, and shear rate, all evaluated at the fiducial radius $r_0$, i.e.,
\begin{align}
\Omega_0(z) &\equiv \Omega(r_0,z)  \,, \\
V_0(z) &\equiv V(r_0,z) \,, \\
S_0(z) &\equiv r_0 \left. \frac{\partial \Omega(r,z)}{\partial r}\right|_{r=r_0} \, .
\end{align}

The operator $\mathcal{D}_0$ is a height-dependent 
generalization of the shearing sheet advection operator 
originally introduced in \citet{1953ApJ...118..106S,1965MNRAS.130..125G}. 
In order to illustrate the action of the advection operator
$\mathcal{D}_0$, Figure \ref{fig_tubes1} shows the effects of 
considering the local bulk flow in Equation~(\ref{eq_local_bulk_flow})
that results from expanding Equation~(\ref{eq_Vrz}) to leading order 
in the radial direction, leaving unaltered the vertical dependence.
In this particular example, we have considered a baroclinic equilibrium
global disk model with a  cylindrical temperature structure, 
which is discussed in detail in Section~\ref{sec_cylindrical_VGSB}.

\subsubsection{Approximation of Momentum and Induction Equations}

Using the approximations above, we arrive at expressions for the momentum 
and induction equations,  Equations~(\ref{momentum_w}) 
and (\ref{eq:induction_D_S}), which are correct to leading order 
in $x/r_0$ and $y/r_0$:
\begin{align}
&\left(\mathcal{D}_0 + \bm{w} \cdot \nabla\right) \bm{w}  
= -2\Omega_0(z)\bm{\hat{z}}\times\bm{w} 
 -S_0(z) w_x\bm{\hat{y}} \nonumber\\
 &- w_z \frac{\partial V(x,z)}{\partial z} \bm{\hat{y}} 
 +  \frac{\nabla P(\rho_h, e_h)}{\rho_h}   
- \frac{\nabla P(\rho, e)}{\rho} 
+ \frac{1}{\rho}\bm{J}\times \bm{B} 
\label{eq_mom_full} \,,
\end{align}
\begin{align}
\mathcal{D}_0{\bm B}=
&
  S_0(z) B_x \hat{\bm y}
+ B_z\frac{\partial V(x,z) }{\partial z}\hat{\bm y}
+ \nabla \times ( {\bm w} \times {\bm B} ) \,.
\label{eq:induction_D_S_0}
\end{align}
Here, all the differential operators are defined in a cartesian 
coordinate system centered at the fiducial radius $r_0$. 

Because we have 
retained the leading order in all the
approximations involving the bulk-flow, the two important flow
properties discussed in Section \ref{sec_eq_motion}
remain unaltered. In the case of an 
inviscid, unmagnetized, barotropic flow
the momentum equation~(\ref{eq_mom_full}) that
results from the local approximation in horizontal planes leads to
\begin{align}
\left(\mathcal{D}_0 + \bm{w} \cdot \nabla\right)  \Gamma = 0\,.
\end{align}
Furthermore, the induction equation~(\ref{eq:induction_D_S_0}) that
results from the local approximation in horizontal planes preserves the 
solenoidal character of the the magnetic field, i.e., 
\begin{align}
\left(\mathcal{D}_0 + \bm{w} \cdot \nabla\right)({\nabla\cdot \bm B})= 0 \,.
\end{align}
This implies that the magnetic flux is frozen into the fluid flow
that results from the local expansion in horizontal planes,
i.e., 
\begin{align}
\left(\mathcal{D}_0 + \bm{w} \cdot \nabla\right) \Phi_{\rm B} = 0 \,.
\end{align}
Therefore, the local approximation in horizontal planes 
leads to equations that still satisfy Kelvin's circulation
theorem and Alfv\'{e}n's frozen in theorem discussed in Section 
\ref{sec_eq_motion}.

\subsection{
Compatibility of the Local Approximation with 
Shearing-periodic Boundary Conditions 
}
\label{sec_compatibility}

\subsubsection{Shearing-periodic Boundary Conditions in the SSB}

All the explicit coordinate dependences in the equations
of motion defining the SSB are contained
in the advection operator
\begin{align}
\mathcal{D}_0^{\rm SSB} &\equiv \frac{\partial}{\partial t} +  x S_0(0) 
\frac{\partial}{\partial y} \label{eq_def_D0_SBB} \,,
\end{align}
which is obtained as a limit of the operator $\mathcal{D}_0$
introduced in Equation (\ref{eq_def_D0}).
The explicit dependence on the coordinate $x$ can be eliminated by the
linear transformation $t' = t$, $x' = x$, $z' = z$, and
\begin{align}
y' =  y- x \, S_0(0) \, t \,.
\end{align}
In the primed coordinate system, the advection operator simply 
becomes $\mathcal{D}_0^{\rm SSB} = \partial_t'$, and the equations 
of motion can be solved by using strictly periodic boundary conditions
in horizontal planes, i.e., 
\begin{align}
f(x',y',z',t') &= f(x'+L_x, y', z',t') \,,
\label{eq:shearing_periodic_x}
\\
f(x',y',z',t') &= f(x', y'+L_y, z',t') \,, \nonumber
\end{align}
and appropriate boundary conditions for the vertical boundaries.
Note that in the original coordinate system, the equations
for the departures from the bulk flow satisfy shearing-periodic
boundary conditions given by
\begin{align}
f(x,y,z,t) &= f(x+L_x, y + S_0(0) L_x t, z,t) \,,\\
f(x,y,z,t) &= f(x, y+L_y, z,t) \,. \nonumber
\end{align}

\subsubsection{Height-dependent Shearing-periodic Boundary Conditions}

Defining an { approximate, radially local} 
set of equations and boundary conditions
for disks with baroclinic equilibria, for which the angular frequency is 
in general a function of height, is more subtle.

The coordinate dependence arising through the
advection operator $\mathcal{D}_0$ in Equation~(\ref{eq_def_D0}) 
can still be removed by defining the
linear transformation  $t' = t$, $x' = x$,  $z' = z$, and
\begin{align}
y' =  y-\left[V_0(z) + S_0(z) \, x \right] t\,.
\end{align}
In this primed coordinate system, the advection operator 
is coordinate-independent, i.e., $\mathcal{D}_0=\partial_t'$.
Therefore, 
if this were the only explicit coordinate dependence then,
in each horizontal plane, it would be enough to consider the 
height-dependent, shearing-periodic boundary conditions given by
\begin{align}
f(x,y,z,t) &= f(x+L_x, y + S_0(z) L_x t, z,t) \,,
\label{eq:shearing_periodic_height_x} \\
f(x,y,z,t) &= f(x, y+L_y, z,t) \,.
 \nonumber
\end{align}
However, the coordinate dependence 
induced by the terms proportional to 
\begin{align}
\frac{\partial V(x,z)}{\partial z} = \frac{\partial V_0(z)}{\partial z} + 
x \frac{\partial S_0(z)}{\partial z} 
 \label{eq_wraping}
\end{align}
on the right-hand sides of Equations~(\ref{eq_mom_full}) and (\ref{eq:induction_D_S_0}) cannot 
be eliminated by the same coordinate transformation that removes
the $x$-dependence in $\mathcal{D}_0$. 
This prevents 
Equation~(\ref{eq_mom_full}) and (\ref{eq:induction_D_S_0}) 
from being solved with the shearing-periodic boundary conditions in
Equation~(\ref{eq:shearing_periodic_height_x}) in a straightforward way.

In what follows we analyze the consequences of proceeding by 
neglecting the term proportional to the coordinate $x$ in 
Equation~(\ref{eq_wraping}). We thus approximate this equation as
\begin{align}
\frac{\partial V(x,z)}{\partial z} 
\simeq \frac{\partial V_0(z) }{\partial z} \,.
\label{eq_wraping_approx}
\end{align}
With the exception of the background hydrostatic profile which will be dealt with in the next section,
this approximation eliminates the explicit coordinate dependence 
on the right-hand sides of Equations~(\ref{eq_mom_full})
and (\ref{eq:induction_D_S_0}), leading to
\begin{align}
&\left(\mathcal{D}_0 + \bm{w} \cdot \nabla\right) \bm{w}  
= -2\Omega_0(z)\bm{\hat{z}}\times\bm{w} 
 -S_0(z) w_x\bm{\hat{y}} \nonumber\\
 &- w_z\frac{\partial V_0(z)}{\partial z}\bm{\hat{y}} 
   +  \frac{\nabla P(\rho_h, e_h)}{\rho_h}   
- \frac{\nabla P(\rho, e)}{\rho}  
+ \frac{1}{\rho}\bm{J}\times \bm{B} 
\label{eq:mom_full_SP_compatible} \,,
\end{align}
\begin{align}
\mathcal{D}_0{\bm B}=
&
  S_0(z) B_x \hat{\bm y}
+ B_z\frac{\partial V_0(z) }{\partial z}\hat{\bm y}
+ \nabla \times ( {\bm w} \times {\bm B} ) \,.
\label{eq:induction_D_S_0_SP_compatible}
\end{align}

Because the only coordinate dependences arise through 
the advection operator $\mathcal{D}_0$, these equations are 
compatible with the height-dependent, shearing-periodic boundary 
conditions Equations~(\ref{eq:shearing_periodic_height_x}). 
By reducing the coordinate dependence down to 
the advection operator $\mathcal{D}_0$, and the 
background equilibrium $\nabla P(\rho_h,e_h)/\rho_h$ 
we have brought these equation closer to compatibility with
with the height-dependent, shearing-periodic boundary 
conditions Equations~(\ref{eq:shearing_periodic_height_x}). 
The background equilibrium will be treated in 
Section~\ref{sec_background_approx},
but first we deal with issues which arise from this step.
The approximation embodied in 
Equation~(\ref{eq_wraping_approx}) does, in general, 
affect the validity of the conservation theorems
discussed in Section \ref{sec_eq_motion}.
We proceed by showing that 
Equations~(\ref{eq:mom_full_SP_compatible})
and (\ref{eq:induction_D_S_0_SP_compatible})
do satisfy  Kelvin's circulation 
theorem and Alfv\'{e}n's frozen-in theorem
when the underlying global disk model
 has a barotropic equilibrium or when we consider axisymmetry.

\subsubsection{Radially local, Vertically Global Hydrodynamic Disk Models}
\label{sec:hydro_models}

It can be seen that equation (\ref{eq:mom_full_SP_compatible}) 
when applied to an inviscid, barotropic, unmagnetized flow,
leads to
\begin{align}
\left(\mathcal{D}_0 + \bm{w} \cdot \nabla\right) \Gamma
= 0 \,.
\end{align}
This means that under the conditions over which Kelvin's circulation
theorem is satisfied, the approximation embodied in Equation~(\ref{eq_wraping_approx}) 
does not lead to spurious 
sources of circulation when considering 
Equation~(\ref{eq_mom_full}) instead of (\ref{eq:mom_full_SP_compatible}).

If the global disk model under consideration 
has baroclinic equilibrium, the circulation $\Gamma$ is no longer 
conserved and thus,  for physical reasons, 
$\left(\mathcal{D}_0 + \bm{w} \cdot \nabla\right) \Gamma$ 
is no longer expected to vanish. However, one should also realize 
that the approximation invoked
in Equation~(\ref{eq_wraping_approx}) leads to a source term 
that contributes spuriously to the evolution 
of the circulation
\begin{align}
\left(\mathcal{D}_0 + \bm{w} \cdot \nabla\right) \Gamma
=\int_{\rm \bf S}  \left[\nabla \times\left(x w_z \frac{\partial S_0(z)}{\partial z} \bm{\hat{y}} \right)\right]
\cdot \bm{dS} + \ldots  \label{eq_circulation_source}
\end{align}
Here, the dots represent the physical sources of circulation
present in fluids which are either viscous or baroclinic.
The spurious source of circulation in Equation~(\ref{eq_circulation_source}) 
vanishes under axisymmetry.
In order to demonstrate this, let us examine the integral 
involved. Using Stoke's theorem, it follows that 
\begin{align}
\int_{\rm \bf S}  \left[\nabla \times\left(x w_z \frac{\partial S_0(z)}{\partial z} \bm{\hat{y}} \right)\right]
\cdot \bm{dS}
  = \oint_{\rm \bf L} \left( x w_z \frac{\partial S_0(z)}{\partial z} \right) 
  \hat{\bm y} \cdot d{\bm l} \,. \,
\end{align}
In axisymmetry, i.e., $\partial_y = 0$, the 
problem reduces to understanding the dynamics of the fluid
in the $(x,z)$ plane. Under this condition, 
the line integral over a closed loop vanishes.

Thus, given a global hydrodynamic disk model, corresponding to either 
barotropic or baroclinic equilibria, Equation~(\ref{eq:mom_full_SP_compatible}) 
can be used to define an associated disk model that is local in horizontal 
planes but global in height. This model can be non-axisymmetric for 
disks with globally barotropic equilibria but should be axisymmetric for 
disks with globally baroclinic equilibria.

The flow of hydrodynamic fluids is constrained by the evolution 
of the potential vorticity as 
governed by Ertel's theorem \citep[see, e.g.][]{1982bsv..book.....P}.
In Appendix~\ref{sec:potvort}, we derive the equations associated 
with the evolution of potential vorticity in the framework of the VGSB and discuss how 
these relate to the Kelvin circulation theorem alluded to in this section.
The details of this also depend on the final approximations needed to make the model compatible with
 shearing periodic radial boundaries in Section~\ref{sec_background_approx}.
Before that, we discuss the analogous issue which occurs with magnetic fields in the induction equation.

\subsubsection{Radially local, Vertically Global MHD Disk Models}
\label{sec:mhd_models}

Let us now consider the implications for the induction equation.
In general, neglecting the term proportional to the 
coordinate $x$ in Equation~(\ref{eq_wraping_approx}) 
leads to an approximated induction equation 
that no longer preserves the solenoidal character of the magnetic field. 
More specifically, Equation~(\ref{eq:induction_D_S_0_SP_compatible}) 
leads to
\begin{align}
\left(\mathcal{D}_0 + \bm{w} \cdot \nabla\right)(\nabla \cdot \bm{B}) = - x\frac{\partial S_0(z)}{\partial z}\frac{\partial B_z}{\partial y} \, .
\label{eq:induction_with_source}
\end{align}
Taken at face value, this implies that Equation~(\ref{eq_wraping_approx}) 
induces spurious generation of magnetic monopoles
that will break flux freezing even in the absence
of dissipation. 
Therefore, in the case of a magnetized fluid, 
while eliminating the explicit $x$-dependence that can not be 
removed by a coordinate transformation allows the induction
equation to be solved with shearing-periodic, height-dependent 
boundary conditions, this approximation, in general, would 
destroy the solenoidal character of the magnetic field. 

However, the spurious evolution of the divergence of the 
magnetic field that results from the approximation 
Equation~(\ref{eq_wraping_approx}) is absent in 
Equation~(\ref{eq:induction_with_source}) if the underlying
global disk model is has a barotropic equilibrium or axisymmetry, i.e., 
$\partial_y = 0$, is considered.

Therefore, given a global disk model, it is possible to develop
models which preserve the solenoidal character of the magnetic field
that are local in horizontal planes
and global in height that are amenable to shearing-periodic
boundary conditions. These models can be non-axisymmetric for 
disks with globally barotropic equilibria but should be axisymmetric for 
disks with globally baroclinic equilibria. 
Under either of these conditions, the 
approximation invoked in
Equation~(\ref{eq_wraping_approx}), which is necessary to employ the
shearing-periodic boundary conditions given by Equation~(\ref{eq:shearing_periodic_height_x}), 
does not lead to spurious
source terms that could affect the evolution of the circulation, 
magnetic flux freezing, or the solenoidal character of the magnetic 
field, i.e.,
\begin{align}
&\left(\mathcal{D}_0 + \bm{w} \cdot \nabla\right)  \Gamma = 0\,, \\
&\left(\mathcal{D}_0 + \bm{w} \cdot \nabla\right)  \Phi_{\rm B}= 0\,,
\end{align}
and
\begin{align}
\left(\mathcal{D}_0 + \bm{w} \cdot \nabla\right)(\nabla \cdot \bm{B}) = 0 \,.
\end{align}

\subsubsection{Approximation of the Background Equilibrium}
\label{sec_background_approx}

The fourth term on the right-hand side of Equation~(\ref{momentum_w}) 
can be dealt with along the lines proposed by  
\citet{1965MNRAS.130..125G} for the hydrostatic background quantities
\begin{align}
\rho_h(r,z) = \rho_{h0}(z) +  \mathrm{O}\left(\frac{x}{r_0}\right) \ , \label{eq_rhohapprox}\\
e_h(r,z) = e_{h0}(z) +  \mathrm{O}\left(\frac{x}{r_0}\right) \ , \label{eq_ethhapprox}
\end{align}
where
$\rho_{h0}(z) \equiv \rho_h (r_0,z)$ and
$e_{h0}(z) \equiv e_h(r_0,z)$ are the leading order terms
associated with the mass and internal energy density 
profiles of the background flow at the fiducial radius\footnote{In the case of a barotropic equation of state, where $e=e(\rho)$ and thus 
$P(\rho,e)=P(\rho)$, only the expansion in Equation~(\ref{eq_rhohapprox}) is 
required, as in \citet[][Section~4]{1965MNRAS.130..125G}.}. 
Within the level of approximation we are working at, we thus have
\begin{align}
\frac{\nabla P(\rho_{h}, e_{h})}{\rho_{h}} = \frac{\nabla P(\rho_{h0}, e_{h0})}
{\rho_{h0}} +  \mathrm{O}\left(\frac{x}{r_0}\right) \,,
\end{align}
where the vertical acceleration induced by the background pressure gradient 
is balanced by gravity at the fiducial radius
\begin{align}
\frac{1}{\rho_{h0}}\frac{\partial P(\rho_{h0}, e_{h0})}{\partial z} =  -\frac{\partial \Phi_0(z)}{\partial z} \bm{\hat{z}} \,,
\end{align}
with $\Phi_0(z) \equiv \Phi(r_0,z)$, the gravitational potential evaluated
at the fiducial radius $r_0$.

{
The set of approximations to the background described 
above can also be considered 
on more formal grounds by introducing a set of dimensionless 
parameters that describe the relative scale of the phenomena 
of interest and the departures from a thin, Keplerian disk.
In Appendix~\ref{sec:nondimensional_hydro_momentum}, we provide 
the details involved in this procedure, emphasizing in particular 
the handling of the terms related to pressure gradients in the 
momentum equation. Our considerations build on, and extend, 
the analysis carried out in \citet{2004A&A...427..855U}.
The analysis suggests the condition $z<\sqrt{x r_0}$ as a limit on the 
height of the best modeled part of the domain.
This criterion also applies to the SSB, as we demonstrate in 
Appendix~\ref{sec:nondimensional_hydro_momentum}.
}

The set of equations that result from considering the approximations
described in this section are compatible
 with the height-dependent, shearing-periodic boundary 
conditions Equations~(\ref{eq:shearing_periodic_height_x}). This motivates 
 introducing the VGSB.

\section{The Vertically Global Shearing Box (VGSB)}
\label{sec_final_eqs}

\subsection{Equations of Motion for the VGSB}
\label{equation_motion_VGSB}

Following the steps outlined in the previous section, 
we arrive at the expressions for the continuity, momentum, 
induction, and energy equations that define the framework 
of the vertically global shearing box (VGSB)
\begin{align}
\mathcal{D}_0 \rho  +  \nabla  \cdot \left( \rho \bm{w} \right)& =  0 \,, \label{eq_vgsb_continuity}
\end{align}
\begin{align}
\left(\mathcal{D}_0 + \bm{w} \cdot \nabla\right) \bm{w}   + w_z\frac{\partial V_0(z)}{\partial z}\bm{\hat{y}}
& \nonumber\\
 = -2\Omega_0(z)\bm{\hat{z}}\times\bm{w} &-S_0(z) w_x \bm{\hat{y}} \nonumber\\
 - \frac{\nabla P}{\rho} -  \frac{\partial \Phi_0(z)}{\partial z} & \bm{\hat{z}}  + \frac{1}{\rho}\bm{J}\times \bm{B} \,,
 \label{eq_vgsb_momentum}
\end{align}
\begin{align}
\left(\mathcal{D}_0 + \bm{w} \cdot \nabla\right) \bm{B}   -B_z \frac{\partial V_0(z)}{\partial z}  \bm{\hat{y}}   & \nonumber\\
=S_0(z) B_x \bm{\hat{y}} +\left(\bm{B}\cdot\nabla\right)\bm{w} -&\bm{B}\left(\nabla \cdot\bm{w}\right)  \,,
\label{eq_vgsb_induction}
\end{align}
\begin{align}
\mathcal{D}_0 e  +  \nabla  \cdot \left( e \bm{w} \right) & = - P \left(\nabla \cdot \bm{w}\right) \,. \label{eq_vgsb_energy}
\end{align}
Here, all the operators are defined in a cartesian coordinate system centered at the fiducial radius $r_0$. 

The only explicit coordinate dependence in the VGSB
arises through the advection operator $\mathcal{D}_{0}$, 
which is linear in the coordinate $x$ for all heights $z$.
This implies that the equations are suitable for being solved with a vertically varying 
shear-periodic $x$-direction boundary condition. The radial and azimuthal boundary 
conditions for mapping a field variable $f$ in a VGSB of size 
$L_x\times L_y\times L_z$ are, respectively,\footnote{The velocity $w_y$ is continuous across the shear-periodic $x$-boundary.}
\begin{align}
f(x,y,z,t) &= f(x+L_x, y + S_0(z) L_x t, z,t) \,,
\label{eq:VGSB_boundaries_x}\\
f(x,y,z,t) &= f(x, y+L_y, z,t) \,.
\label{eq:VGSB_boundaries_y}
\end{align}
In order to completely define the problem,  
appropriate boundary conditions in the vertical direction
must be specified.

The set of Equations~(\ref{eq_vgsb_continuity})--(\ref{eq_vgsb_energy}), 
together with the boundary conditions
Equations~(\ref{eq:VGSB_boundaries_x})--(\ref{eq:VGSB_boundaries_y})
lead to 
{ approximate radially local, vertically global disk models}
if the underlying global disk model has a barotropic equilibrium or if we assume axisymmetry.

It is important to emphasize that the velocity $\bm{w}$ is the departure from the local 
approximation of the bulk flow and thus the total fluid velocity 
$\bm{v}$ in the VGSB is
\begin{align}
 \bm{v}   = \left[V_0(z) + S_0(z)x\right]\bm{\hat{y}} + \bm{w} \,, \label{eq_vgsb_fluctuation_transform}
\end{align}
where
\begin{align}
V_0(z) = r_0 [\Omega_0(z)-\Omega_0(0)] \,, \label{V_0}
\end{align}
and the local shear rate $S_0(z)$ is related to the generalization of the 
height-independent  $q$-parameter in the SSB:
\begin{align}
q_{0}(z) \equiv -\frac{S_0(z)}{\Omega_0(0)} \, .
\end{align}
At the midplane, i.e., $z=0$, the flow velocity $V(x,0) = S_0(0)\,x = -q_{0}(0) \Omega_0(0) \,x$ 
is the same as the steady state flow velocity in the SSB. However, 
in global disk models where the shear rate of the flow decreases to zero at high altitude, i.e.,
$S_0(z) \simeq 0$ and, therefore, $V_0(z) \simeq - r_0 \Omega_0(0)$ for $z\gg r_0$.
This is just the reflex motion induced by the rotating frame. This shows that in the limit $z\gg r_0$,
Equations~(\ref{eq_vgsb_continuity})--(\ref{eq_vgsb_energy})
reduce to the MHD equations in the absence of a gravitational field.

\subsection{Connecting the VGSB to the SSB}

\label{sec_polytropic_VGSB}

In order to connect the SSB to the VGSB, with its generalization to baroclinic disk equilibria,
it is useful to recall some results that apply to 
barotropic hydrostatic equilibria. If the pressure is $P=P(\rho)$
it follows that:
\begin{itemize}
\item[(i)] The angular frequency $\Omega$ is independent of $z$. 
This can be derived by taking the curl of the momentum 
equation~(\ref{eq_mom_barotropic}), which leads to
\begin{align}
r \frac{\partial \Omega^2}{\partial z} =  0 \,,
\label{eq_TaylorProudman}
\end{align}
implying that barotropic equilibria rotate on cylinders,\footnote{
This is often referred to as the Taylor-Proudman theorem, which is commonly credited 
to \citet{1917RSPSA..93...99T,1916RSPSA..92..408P}.
The same principle that we reference here, that barotropes rotate on cylinders,
 is also referred to in the literature 
 as the Poincar\'e-Wavre therom \citep[][Section 3.1.2]{2000stro.book.....T}
and the von~Zeipel condition \citep{1924MNRAS..84..665V}.} 
i.e.,~$\Omega \equiv \Omega(r)$.
\item[(ii)] The generalized gravito-thermal potential $\Theta$ is 
independent of $z$.
This follows from the vertical component of Equation~$(\ref{eq_TaylorProudman})$, 
because the bulk flow velocity has only an an azimuthal component. This implies that
\begin{align}
\frac{\partial \Theta}{\partial z} = 0 \,,
\label{eq_mom_barotropic_z}
\end{align}
and thus $\Theta \equiv \Theta(r)$, as stated.
\item[({iii})] The angular frequency can be obtained from 
the generalized gravito-thermal potential using the radial component of 
Equation~(\ref{eq_mom_barotropic}) as
\begin{align}
r \Omega^2 = \frac{\partial \Theta}{\partial r} \,.
\label{eq_mom_barotropic_r}
\end{align}
\end{itemize}

Let us now focus our attention on a global isothermal disk 
in the gravitational potential of a point source of mass 
$M$, i.e.,
\begin{align}
\Phi(r,z) = \frac{-GM}{\sqrt{r^2+z^2}} \,,
\end{align}
with $G$ the gravitational constant. The 
pressure is $P=\rho c_{{\rm s} 0}^2$,  with constant sound speed $c_{{\rm s} 0}$.
Adopting a power-law  density dependence with radius in the midplane, 
tha hydrostatic balance in the vertical direction implies that the density is
\begin{align}
\rho(r,z) &= \rho_0 \left(\frac{r}{r_0}\right)^p
\exp\left[-\frac{v_{\rm K}^2(r)}{c_{{\rm s} 0}^2}\left(1-\frac{1}{\sqrt{1+(z/r)^2}}\right)\right]\,,
\label{eq_density_vgsb_to_ssb}
\end{align}
where the constant $p\le 0$ and we defined the Keplerian speed
\begin{align}
 v_{\rm K}(r) = \sqrt{\frac{GM}{r}} \,.
\end{align}

In this case, 
the enthalpy is simply $h=c_{{\rm s} 0}^2 \ln\rho$, and thus the generalized 
gravito-thermal potential $\Theta = 
\Phi+h$, is 
\begin{align}
\Theta = c_{{\rm s} 0}^2 \ln\left[\rho_0 \left(\frac{r}{r_0}\right)^{\!p} \,\right] - v_{\rm K}^2(r) \,.
\end{align}
The angular frequency can be obtained from 
Equation~(\ref{eq_mom_barotropic_r}) as
\begin{align}
\Omega = \Omega_{\rm K}(r) \sqrt{1 + p\,\frac{c_{{\rm s} 0}^2}{v_{\rm K}^2(r)}} \,,
\label{eq_omega_vgsb_to_ssb}
\end{align}
where the Keplerian frequency is 
\begin{align}
 \Omega_{\rm K}(r) = \frac{v_{\rm K}(r)}{r} = \sqrt{\frac{GM}{r^3}} \,.
\end{align}
The generalized gravito-thermal potential $\Theta$ and the angular frequency $\Omega$ are
both independent of height, as expected.

Expanding Equations~(\ref{eq_density_vgsb_to_ssb}) and (\ref{eq_omega_vgsb_to_ssb}) 
in radius around $r_0$, we obtain
expressions for the local values the angular frequency $\Omega_0$ and the shear rate $S_0$
that can be used, in the framework of the VGSB, to study isothermal disks which are not 
necessarily thin compared to the local radius $r_0$, i.e., 
\begin{align}
\rho_0(z) & = \rho_0 \exp\left[- 
\frac{{v_{{\rm K} 0}^2}}{{c_{{\rm s} 0}^2}}
\left(1-\frac{1}{\sqrt{1+(z/r_0)^2}}\right)\right]\,, \\
\Omega_0 & = \Omega_{{\rm K} 0} \sqrt{1+p\frac{c_{{\rm s} 0} ^2}{v_{{\rm K} 0}^{2}}} \,,\\
S_0 &=-\frac{3}{2}\Omega_{{\rm K} 0} 
\left(1+ \frac{2p}{3} \frac{c_{{\rm s} 0}^{2}}{v_{{\rm K} 0}^2} \right)
 \left( 1+ p  \frac{c_{{\rm s} 0}^2}{v_{{\rm K} 0}^{2}} \right)^{-1/2}  \,,
\end{align}
where we have defined $v_{{\rm K} 0} \equiv v_{\rm K}(r_0)$ and $\Omega_{{\rm K} 0} \equiv \Omega_{{\rm K} 0}(r_0)$.

These expressions are useful to show that in the limit of a cold, thin disk, i.e.,
$c_{{\rm s} 0} \ll v_{{\rm K} 0}$, the density profile becomes
$\rho_0(z) = \exp[-(z/H_0)^2/2]$, with $H_0/r_0 = c_{{\rm s} 0}/v_{{\rm K} 0}$,
and the angular frequency and shear rate become
$\Omega_0 = \Omega_{{\rm K} 0}$,  $S_0 = -3/2  \Omega_{{\rm K} 0}$, respectively,
and Equations~(\ref{eq_vgsb_continuity})--(\ref{eq_vgsb_energy}) reduce to 
the equations for the SSB.

\begin{figure*}
\begin{center}
\includegraphics[width=5.9cm]{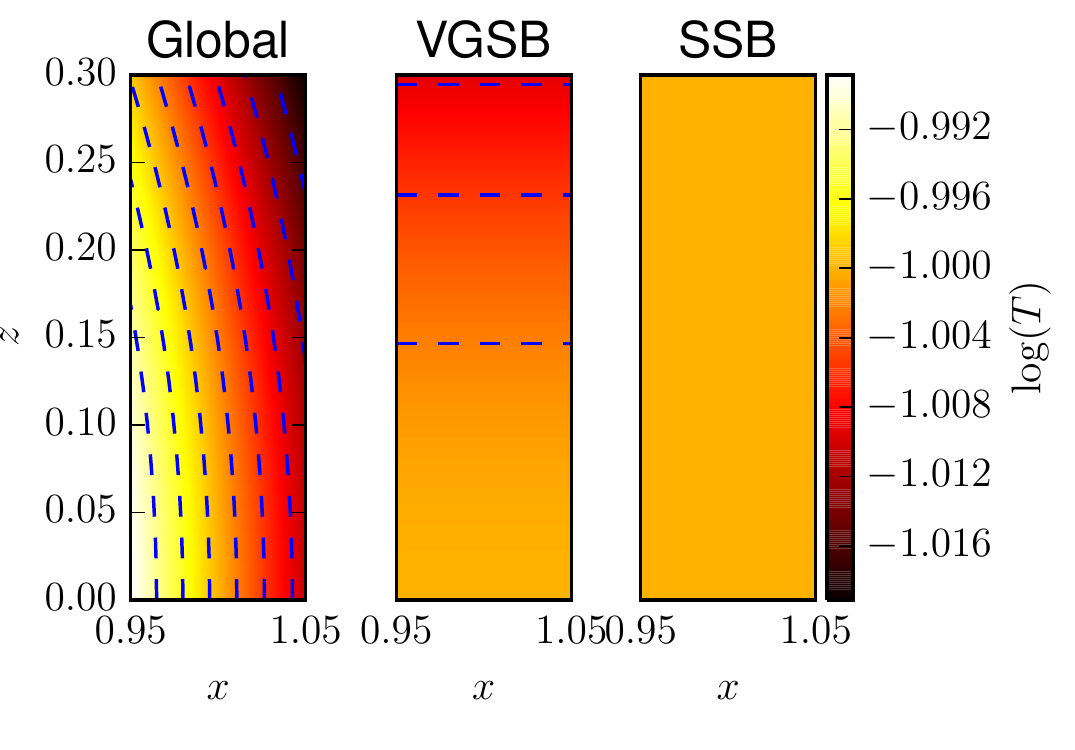}
\includegraphics[width=5.9cm]{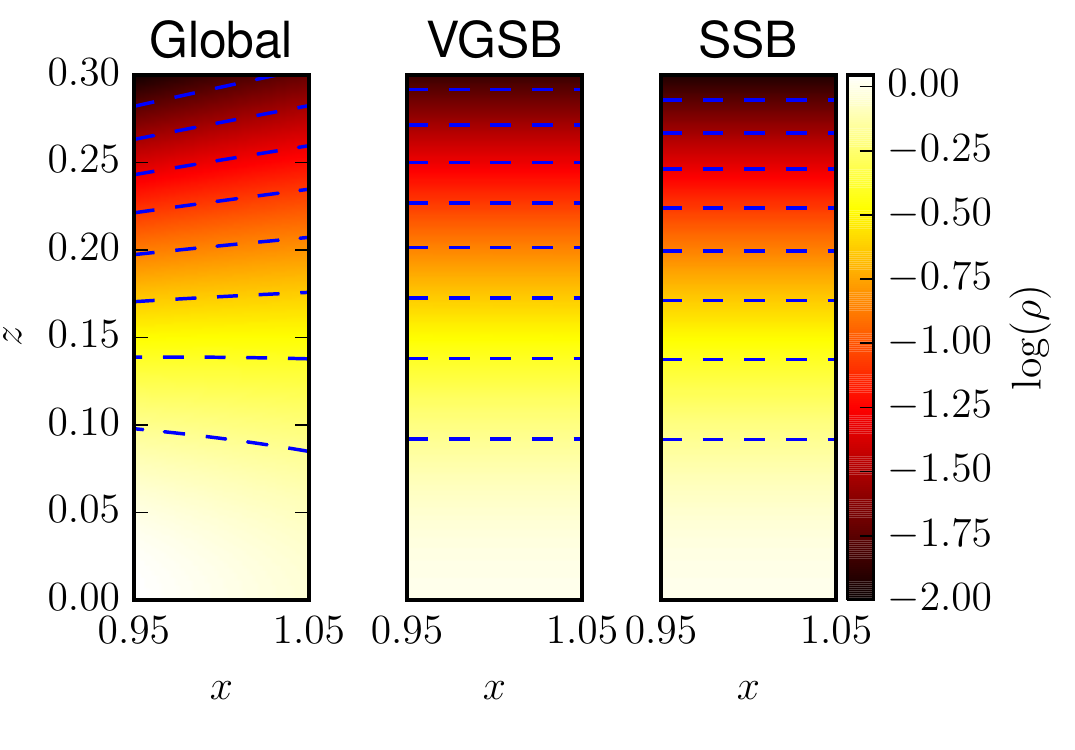}
\includegraphics[width=5.9cm]{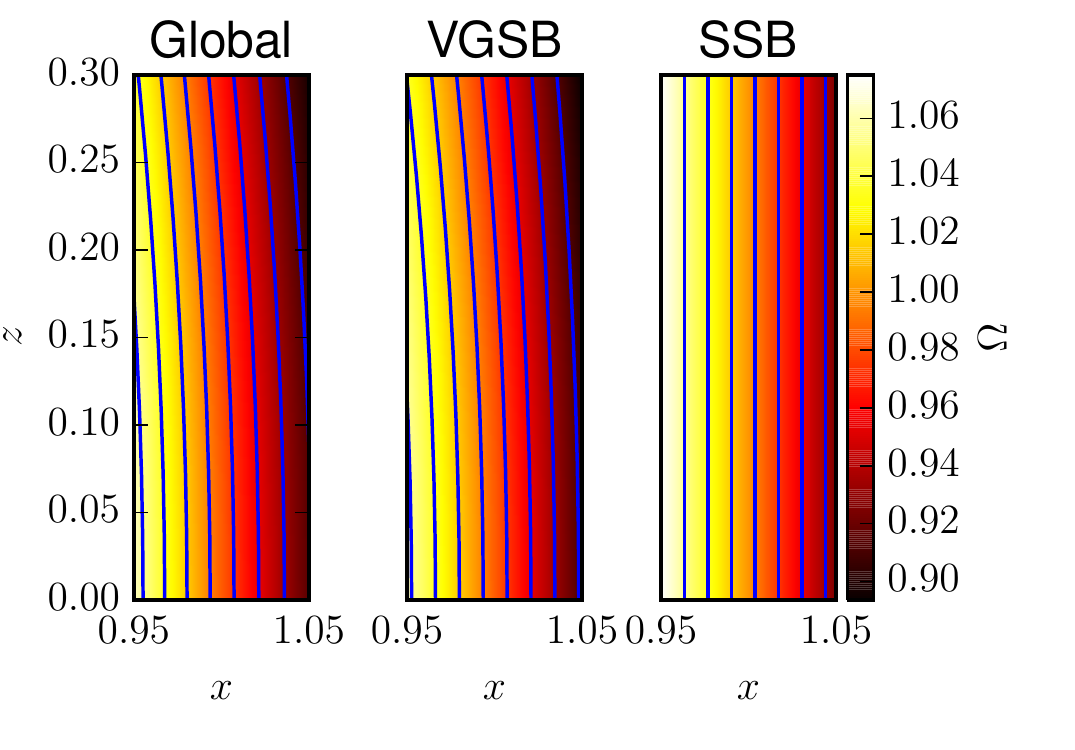}
\end{center}
\caption{A global hydrostatic equilibrium with spherical temperature structure and local approximations. The analytical forms for the global and VGSB 
equilibrium are given in Section~\ref{sec_spherical_VGSB}, with $\mu=3$ and $\nu=97$, yielding an aspect ratio of $0.1$. 
The three panels show, from left to right, the logarithm of temperature, logarithm of pressure, and angular velocity. 
{\em Global:} exact global equilibrium. {\em VGSB:} vertically global, horizontally local approximation. 
{\em SSB:} standard shearing box, horizontally and vertically local.
In line with the analysis in Appendix~\ref{sec:nondimensional_hydro_momentum}
these plots show only the regions $z<\sqrt{x r_0}$.}
\label{fig_localhydro}
\end{figure*}

\subsection{Examples of VGSB Models for Global Equilibria}
We now consider two families of baroclinic, global disk equilibria that 
are characterized by angular frequencies and shear rates that depend on height, 
and are thus impossible to study within the standard shearing box approximation.
The  VGSB framework, embodied in Equations~(\ref{eq_vgsb_continuity})--(\ref{eq_vgsb_induction}), 
together with their associated boundary conditions, can be used to produce 
axisymmetric, vertically global, 
horizontally local models for these astrophysical disks. 
After these examples, we summarize the new aspects of the baroclinic VGSB.

\subsubsection{VGSB for Global Disk Models with Cylindrical Temperature Structure}
\label{sec_cylindrical_VGSB}

There is a family of disk models which are isothermal in the vertical direction
at any fiducial radius. These have been used, for example, by \citet{2013MNRAS.435.2610N}.
The corresponding global hydrostatic equilibrium 
configuration is given by
\begin{align}
T(r) &\equiv T_0\left(\frac{r}{r_0}\right)^q \,, \\
\rho(r,z) &= \rho_0 \left(\frac{r}{r_0}\right)^p 
\exp\left[-\frac{v_{\rm K} ^{2}}{c_{\rm s}^2}\left(1-\frac{1}{\sqrt{1+(z/r)^2}}\right)\right]\,, 
\nonumber \\ \\
\Omega(r,z) &=   \Omega_{\rm K}  \sqrt{1+(p+q) \frac{c_{\rm s}^2}{v_{\rm K} ^{2}} + q \left(1-\frac{1}{\sqrt{1 + (z/r)^2}}\right)} \, ,
\label{omega_T_cyl}
\end{align}
where $c_{\rm s}$ stands for the sound speed, assuming an ideal gas,
\begin{align}
c_{\rm s}^2(r) = c_{{\rm s} 0} ^2 \left(\frac{r}{r_0}\right)^q \,.
\end{align}
The local expansions involved in the VGSB
for this global baroclinic equilibrium disk model are
\begin{align}
\Omega_0(z) & = \Omega_{{\rm K} 0} \sqrt{1+(p+q)\frac{c_{{\rm s} 0} ^2}{v_{{\rm K} 0}^{2}} + q\left(1 -\frac{1}{\sqrt{1 +(z/r_0)^2}}\right)} \,, \label{eq_cylvgsbomega}
\end{align}

\begin{align}
&S_0(z) =-\frac{3}{2}\Omega_{{\rm K} 0} \nonumber \\
&\times \left[ 1+ \frac{(2-q)(p+q)}{3} \frac{c_{{\rm s} 0} ^{2}}{v_{{\rm K} 0}^2} + q \left(1- \frac{1+(2/3)(z/r_0)^2}{[1+(z/r_0)^2]^{3/2}}\right)\right] 
\nonumber \\
&\times \left[ 1+ (p+q) \frac{c_{{\rm s} 0} ^2}{v_{{\rm K} 0}^{2}} + q\left(1 - \frac{1}{\sqrt{1 + (z/r_0)^2}}\right)\right]^{-1/2}  \,, \label{eq_cylvgsbS}
\end{align}
with $v_{{\rm K} 0} = v_{\rm K}(r_0)$ and $\Omega_{{\rm K} 0} = \Omega_{{\rm K} 0}(r_0)$.

Note that all the explicit $z$--dependences in $\Omega_0(z)$ and $S_0(z)$
arise only when $q\ne0$. This is in agreement with the isothermal case, corresponding
to $q=0$, in which the angular frequency is independent of height.
It can also be seen from inspection of these equations that in 
the thin disk limit, i.e., $c_{{\rm s} 0} /v_{{\rm K} 0} \rightarrow 0$,
the VGSB reduces to the SSB
with $\Omega_{0}(0)=\Omega_{{\rm K} 0}$, $V_0(z)=0$, and 
$S_0(z) = -(3/2)\Omega_{{\rm K} 0}$, provided that 
$z\ll r_0$, even if the temperature depends on radius,
i.e., $q\ne 0$.
This is what ultimately justifies the use of a 
collection of SSBs to produce local 
barotropic equilibrium disk models, with height-independent angular frequency
at different radii, even though the underlying global disk 
model has a baroclinic equilibrium structure.

\subsubsection{VGSB for Global Disk Models with Spherical Temperature Structure}
\label{sec_spherical_VGSB}

A spherical temperature dependence of the form
\begin{align}
T(r,z)   &\equiv T_0 \frac{r_{0}}{\sqrt{r^2+z^2}} \,,
\end{align}
leads to a global hydrostatic disk configuration given by
\begin{align}
\rho(r,z)&= \rho_0 \left(\frac{r}{\sqrt{r^2+z^2}}\right)^\nu \left(\frac{\sqrt{r^2+z^2}}{r_0}\right)^{1-\mu} \,,\\
\Omega(r,z) & = \sqrt{\nu} \, \frac{c_{{\rm s} 0} }{r} \left(\frac{\sqrt{r^2+z^2}}{r_0}\right)^{-1/2} \,.
\end{align}
Here, the power-law coefficients $\nu$ and $\mu$ are related via
\begin{align}
\nu+\mu & = \frac{v_{{\rm K} 0}^{2}}{c_{{\rm s} 0} ^2} \,,
\end{align}
and the sound speed is
\begin{align}
c_{\rm s}^2 (r,z) =c_{{\rm s} 0}^2  \frac{r_{0}}{\sqrt{r^2+z^2}} \,,
\end{align}
where $c_{{\rm s} 0}  = c_{\rm s}(r_0,z)$ is the sound speed at the midplane at $r=r_0$.
Such models have been used, for example, by \citet{2014ApJ...784..121S}.

The expansion of these expressions leads to the VGSB model 
corresponding to a global disk with a temperature structure
varying in spherical shells. The local angular frequency and shear
rate are, respectively,
\begin{align}
\Omega_0(z) &= \Omega_{{\rm K} 0} \sqrt{ \frac{\nu}{\nu+\mu}} \left[1+\left(\frac{z}{r_0}\right)^2\right]^{-1/4} \,, \\
S_0(z) & = -  \Omega_{{\rm K} 0} \sqrt{ \frac{\nu}{\nu+\mu}} \nonumber\\
 &\qquad \times\left[\frac{3}{2}+\left(\frac{z}{r_0}\right)^2\right]\left[1+\left(\frac{z}{r_0}\right)^2\right]^{-5/4} \,.
\end{align} 

In the limiting case in which $\mu = 0$; the angular frequency and the shear 
rate take the Keplerian values at the midplane, i.e., 
$\Omega_0(0) = \Omega_{{\rm K} 0}$ and $S_0(0) = -(3/2) \Omega_{{\rm K} 0}$,
and the parameter $\nu = v_{{\rm K} 0}^2/c_{{\rm s} 0}^2$ alone determines 
the disk thickness.
 
 \subsubsection{New Aspects of the VGSB}
\label{sec_accomplishment}

The approach that we followed in deriving the equations defining the
framework of the VGSB is similar in spirit
to the one employed in the derivation of the SSB.
This consists of expanding the 
steady sate bulk flow, retaining only the leading order terms in the
ratio $x/r_0$. The significant difference resides in that we have
avoided making any expansion in the vertical direction. 
This allows us to retain the full height-dependence of the angular frequency.
This is essential when dealing with disks in which the equilibrium structure 
is described by a baroclinic equation of state, since their angular
frequency is in general a function of height.

In order to illustrate the new aspects that are open to examination
by retaining the full height-dependence in Equations~(\ref{eq_omegarz_exp}),
(\ref{eq_rhohapprox}), and (\ref{eq_ethhapprox}), let us consider a
baroclinic equilibrium global disk model, in which the temperature is a
function of the spherical radius, as in
\citet{2014ApJ...784..121S}. The four panels in
Figure~\ref{fig_localhydro} compare the temperature, density,
pressure, and angular frequency corresponding to the global
equilibrium of the disk model (labeled ``Global''), the local
equilibrium defining the VGSB, and
local equilibrium that would result in the SSB.

The  VGSB provides a local representation of a global 
baroclinic equilibrium disk model, capturing effects that the 
SSB would not be able to account for.
Because the lowest order discarded in Equation~(\ref{eq_omegarz_exp}) is
$\mathrm{O}(x^2/r_0^2)$, the radial dependence of the angular frequency
is correct, to linear order in radius, for all heights.\footnote{
 Note that the radially local expansion removes the radial variation of temperature, resulting in 
a barotropic pressure-temperature relationship in the VGSB equilibrium, while the vertical shear of the global 
baroclinic equilibrium is retained.}
Moreover,
the  VGSB retains the correct global vertical gradients in temperature,
density, and pressure, while neglecting the local radial variation of
the isopycnic, isothermal, isobaric surfaces.
This is in sharp contrast with the local equilibrium
involved in the SSB framework that
is unable to capture the variation of the angular frequency 
with height, preventing its use when modeling disks with a baroclinic 
equilibrium structure. 
 Note also that, for the case under consideration,
the equilibrium density and pressure 
profiles involved in the standard isothermal shearing box 
are less accurate than the ones associated with the VGSB.

In the next two sections, we illustrate how the VGSB framework
relates to, and extends, previous treatments of two disk instabilities 
that are relevant for a wide variety of astrophysical disks;
namely the VSI, also known as the Goldreich-Schubert-Fricke (GSF) 
instability \citep{1967ApJ...150..571G,1968ZA.....68..317F},
and the MRI \citep{1991ApJ...376..214B}.
These instabilities will be the subject of future, more detailed, work.

\section{VSI in the VGSB}
\label{sec_vsi}

Unmagnetized disks with shear profiles that depend on height have 
been long suspected to be unstable to various instabilities that 
feed off this angular frequency gradient. The pioneering studies of
\citet{1967ApJ...150..571G} and \citet{1968ZA.....68..317F}
invoked local approximations in both
radius and height, capturing the essence of these instabilities 
but leaving open questions about their global behavior. 
These instabilities have been studied locally, and in conjunction with 
magnetorotational instability, in the context of accretion disks. 
It has been suggested that they can play a role in the low-conductivity 
regime characterizing protoplanetary disks
\citep{1998MNRAS.294..399U,2002A&A...391..781R,2003A&A...404..397U,2004A&A...426..755A}.
The global, nonlinear evolution of these instabilities has ben recently 
studied in \citet{2013MNRAS.435.2610N} by performing numerical simulations.
This work includes an extension of the original local analysis, 
by considering
the effects of compressibility and also an approximate, vertically global 
linear mode analysis. In this section, we show that the VGSB formalism 
recovers the local dispersion relation found in \citet{2013MNRAS.435.2610N}
and can be used to address their vertically global mode analysis without 
invoking the approximations related to compressibility 
considered by the authors.\footnote{After a preprint of this paper was first posted on the arXiv, several other works have treated the VSI, including \citet{2014A&A...572A..77S, 2015MNRAS.450...21B, 2015arXiv150502163L}, and \citet{2015arXiv150501892U}. Notably \citet{2015arXiv150502163L} discuss the effects of fully removing the background radial pressure gradient, as done in this section. 
}

There are general considerations which are common to both local and 
global approaches
in height.  Let us first write the VGSB equations 
for an unmagnetized disk for which the local angular frequency and shear rate 
derive from a global disk model with temperature $T(r)$
dependent on the cylindrical radius,
as described in Section~\ref{sec_cylindrical_VGSB}. Following 
\citet{2013MNRAS.435.2610N}, we make the change of variables $\Pi\equiv\log\rho$ so that 
Equations~(\ref{eq_vgsb_continuity})--(\ref{eq_vgsb_energy}) become
\begin{align}
&\mathcal{D}_0 \Pi  + \bm{w}\cdot\nabla\Pi + \nabla  \cdot \bm{w}  =  0 \,,
 \label{eq_vsi_general_cont} \\
&\left(\mathcal{D}_0 + \bm{w} \cdot \nabla\right) \bm{w}   + w_z\frac{\partial V_0(z)}{\partial z}\bm{\hat{y}}
=   \nonumber\\
&\qquad -2\Omega_0(z)\bm{\hat{z}}\times\bm{w} -S_0(z) w_x \bm{\hat{y}} 
 - \frac{\nabla P}{\rho} -  \frac{\partial \Phi_0(z)}{\partial z}  \bm{\hat{z}} \,.
 \label{eq_vsi_general_mom}
\end{align}

In order to derive the equations for the linear mode analysis that lead to the VSI
starting from Equations~(\ref{eq_vsi_general_cont})~and~(\ref{eq_vsi_general_mom}), we proceed as follows:
(\emph{i}) we write the density variable as $\Pi = \Pi_h + \Pi_f$, i.e., the sum of its hydrostatic 
value and a fluctuation over this background, and the pressure as $P=c_{\rm s}^2(r_0,z)\rho$.
(\emph{ii}) We use the scales of length and time provided by $r_0$ and $\Omega_0(0)^{-1}$
to define the dimensionless variables
$\bm{x}/r_0\rightarrow \bm{x} $ and $t \Omega_0(0)\rightarrow t$, 
so that $S_0(z)/\Omega_0(0)\rightarrow S_0(z)$ is the dimensionless shear rate,
$c_{{\rm s}0}/r_0 \Omega_0(0) \rightarrow c_{\rm s}$ is the dimensionless sound speed, etc.
(\emph{iii}) We assume that all the perturbations are small and axisymmetric. 
(\emph{iv}) We focus on radial scales that are small compared to the fiducial radius $r_0$
and take the Fourier transform of the set of equations~(\ref{eq_vsi_general_cont})~and~(\ref{eq_vsi_general_mom}), 
which reduces to making the substitution $f(x,z) \rightarrow \tilde{f}(k_x,z)\exp(\sigma t + ik_x x)$,
for all flow variables, $\Pi_f$, $w_x$, $w_y$, and $w_z$, where the tilde denotes Fourier amplitudes. 
This procedure leads to the following set of dimensionless equations
\begin{align}
\sigma  \tilde{\Pi}_f  &=  - \tilde{w}_z \partial_z \Pi_h - i k_x \tilde{w}_x - \partial_z \tilde{w}_z  \,,
\label{eq_VSI_general_Pi} \\
\sigma \tilde{w}_x  &= 2 \Omega_0(z)\tilde{w}_y - i k_x c_{\rm s}^2 \tilde{\Pi}_f \,,\\
\sigma \tilde{w}_y  &= - \left[2 \Omega_0(z)+ S_0(z) \right] \tilde{w}_x - \tilde{w}_z \partial_z V_0(z)\,,\\
\sigma \tilde{w}_z  &= - c_{{\rm s}0}^2 \partial_z \tilde{\Pi}_f \, .
\label{eq_VSI_general_wz}
\end{align}

\label{sec_vsigbl}

\begin{figure}
\plotone{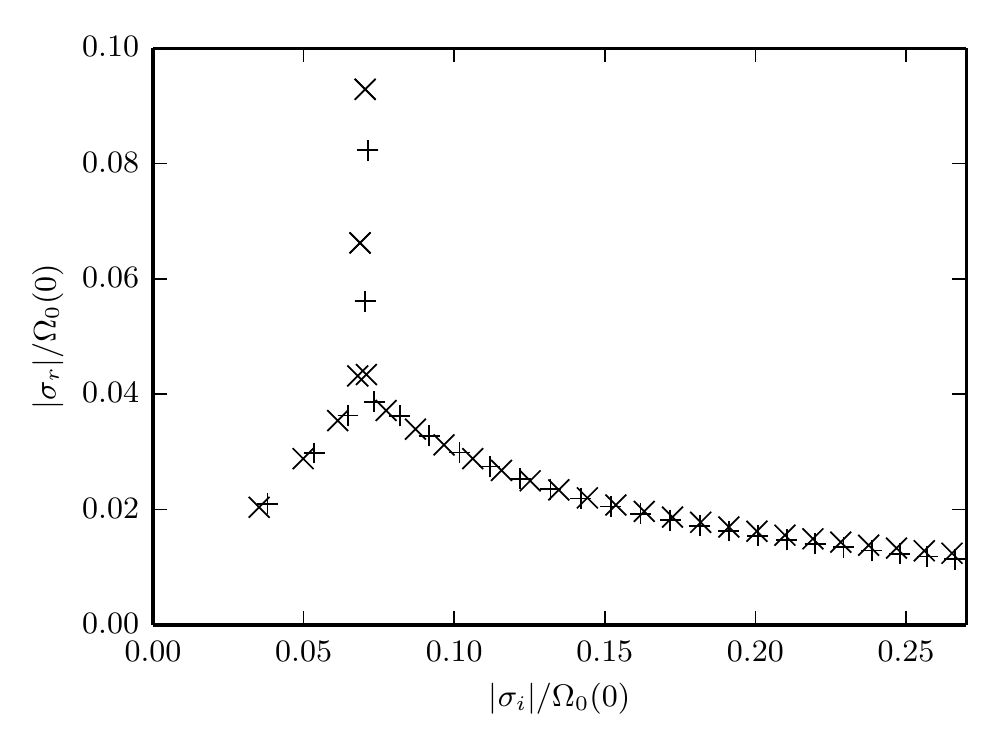}
\caption{Comparison of the eigenvalues for the \citet{2013MNRAS.435.2610N} approximation (their equation~39 \textsuperscript{\ref{foot-nelson}})
 and the fully compressible
analysis given here for the parameters as listed in \citet{2013MNRAS.435.2610N} 
(in our scaling $k_x=+200\pi$, $p=-1.5$, $q=-1$, $c_{\rm s0}/v_{\rm K}=0.05$,$-5\le z/H_{\rm c} \le 5$).
To convert the values here to those scaling of \citet{2013MNRAS.435.2610N}  multiply by $2\pi$.
Symbols: $+$ denotes our result, $\times$ denotes result calculated following \citet{2013MNRAS.435.2610N}.
The main effect of including compressibility is to reduce the growth rate of the surface modes.
}
\label{fig3}
\end{figure}

\begin{figure*}
\includegraphics[width=\textwidth]{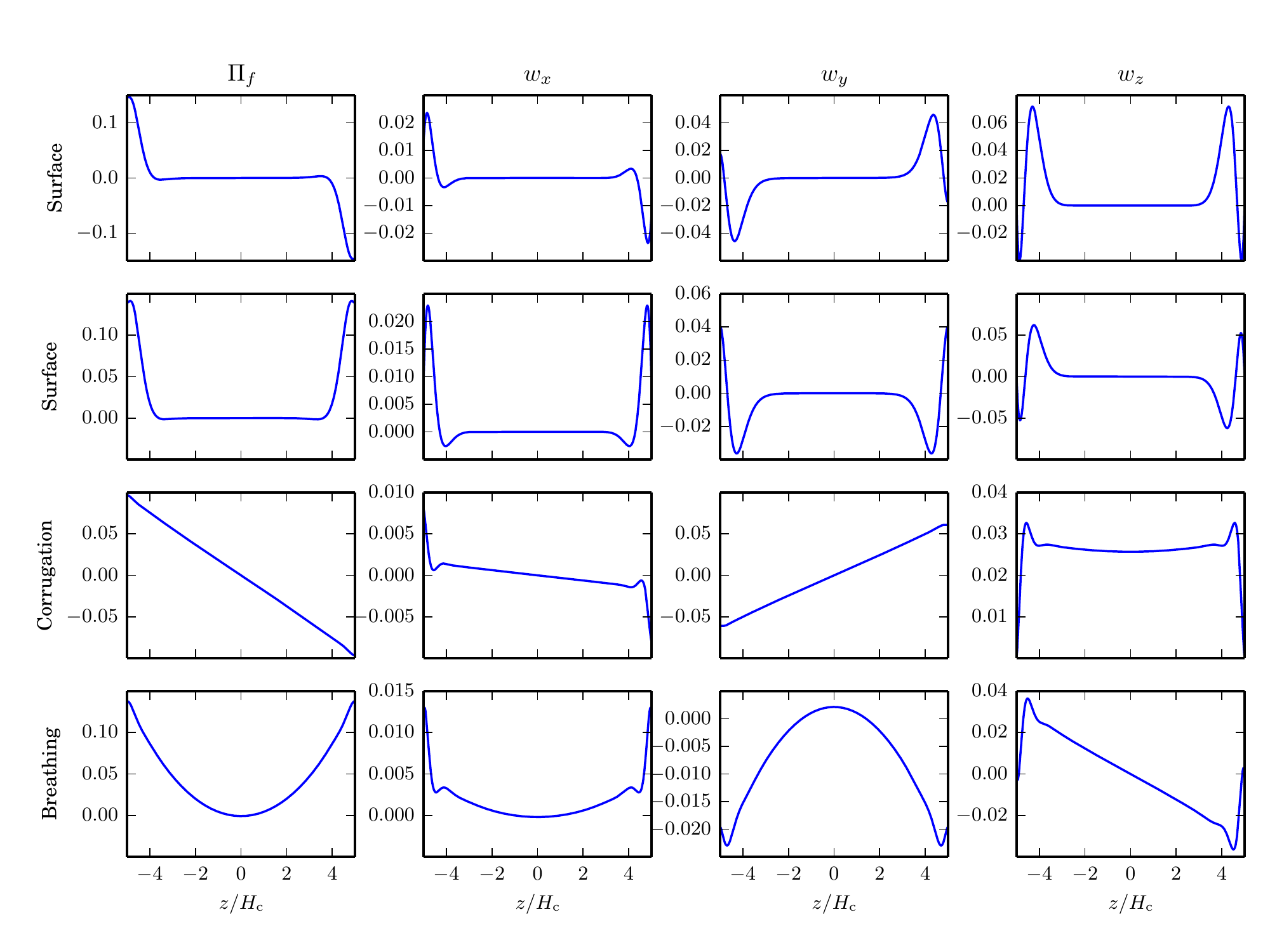}
\caption{Four example eigenfunctions, transformed to real space, of the VSI, with normalization as given in text. 
{\em Top Two Rows:} the two degenerate fastest growing surface modes with 
degenerate eigenvalues, both modes have
$\sigma \approx 0.0823 \pm 0.0713 i$. 
{\em Third Row:} fundamental corrugation mode with
$\sigma \approx 0.0210 \pm 0.0380 i$.
{\em Fourth Row:} fundamental breathing mode with
$\sigma \approx 0.0298 \pm 0.0535 i$. 
}
\label{fig4}
\end{figure*}

\subsection{Vertically Local Analysis of VSI}

Let us first show that the equations for the VGSB lead to the dispersion 
relation derived by \citet[][section 6.1]{2013MNRAS.435.2610N}, which is a 
generalization of the dispersion relation derived originally by GSF
\citep{1967ApJ...150..571G,1968ZA.....68..317F}, who considered an incompressible gas.
In order to do this, we restrict the analysis to the vertical location about the 
height $z=z_0$, focusing on scales that are small in both the radial and vertical direction.
Note that, in this case, the convenient time scale to define dimensionless variables is $\Omega_0(z_0)^{-1}$.
Taking the Fourier transform of the set of equations~(\ref{eq_vsi_general_cont})~and~(\ref{eq_vsi_general_mom}) 
reduces to making the substitution $f(x,z) \rightarrow 
\tilde{f}(k_x,k_z)\exp(\sigma t + ik_x x + ik_z z)$, so that
\begin{align}
\sigma  \tilde{\Pi}_f  &=  \tilde{w}_z \frac{g}{c_{{\rm s}0}^2} - i k_x \tilde{w}_x - i k_{z} \tilde{w}_z  \,,
\label{eq_VSI_general_dimensionless_Pi}
\\
\sigma \tilde{w}_x  &= 2 \Omega_0 \tilde{w}_y - i k_x c_{{\rm s}0}^2 \tilde{\Pi}_f \,,\\
\sigma \tilde{w}_y  &= - \left(2 \Omega_0+ S_0 \right) \tilde{w}_x - \tilde{w}_z \partial_z V_0 \,,\\
\sigma \tilde{w}_z  &= - i k_{z} c_{{\rm s}0}^2 \tilde{\Pi}_f \, ,
\label{eq_VSI_general_dimensionless_wz}
\end{align}
where the background quantities are understood to be evaluated at $z_0$
and we have used that $\partial_{z} \Pi_h = -g/c_{{\rm s}0}^2$,
where $g$ is the local value of the gravitational accelleration.

The characteristic polynomial of the homogeneous system 
of equations~(\ref{eq_VSI_general_dimensionless_Pi})--(\ref{eq_VSI_general_dimensionless_wz})
yields the dispersion relation in terms
of dimensionless variables
\begin{align}
 \sigma^4  &+ \sigma^2\left[  c_{{\rm s}0}^2( k_x^2  + k_z^2)  +ig k_z + 2(S_0 + 2\Omega_0)\Omega_0  \right]
 \nonumber\\
 & + 2(S_0+2\Omega_0)\Omega_0 ( c_{{\rm s}0}^2 k_z^2 + i k_z g ) -2 \Omega_0   c_{{\rm s}0}^2k_x k_z  \partial_{z} V_0
    = 0 \,.
\end{align}
This result is equivalent to the equation one before Equation~(32) in \citet[][]{2013MNRAS.435.2610N}.
Therefore, the dispersion relation of the VSI present in global 
disk models with height-dependent angular frequencies is correctly obtained
by the VGSB framework.

\subsection{Vertically Global Analysis of VSI}

In this section we perform a vertically global analysis of isothermal, axisymmetric 
unmagnetized perturbations by solving the coupled set of ordinary differential 
equations~(\ref{eq_VSI_general_Pi})--(\ref{eq_VSI_general_wz}). For a fixed radial 
wavenumber $k_{x}\gg1$, the solution 
to the eigenvalue problem defined by these equations yields a set of eigenfunctions
$\tilde{\Pi}_f$, $ \tilde{w}_x$, $\tilde{w}_y$, and $\tilde{w}_z$ associated with the
eigenvalues $\sigma$. We are particularly interested in finding the modes with
growing amplitude, i.e., where the real part of the eigenvalue $\sigma_r > 0$.

We compare the eigenvalues for a given set of parameters to those found 
through the approximated eigenproblem derived by \citet{2013MNRAS.435.2610N},  which invokes
additional approximations. Reasonable agreement is found,
as shown in Figure~\ref{fig3}, with the main difference being that 
the surface modes show slower growth when full compressibility is retained.

We illustrate the  eigenmodes present in a disk model with cylindrical 
temperature structure $T(r)$, as presented in Section~\ref{sec_cylindrical_VGSB}, 
in Figures~\ref{fig3} and \ref{fig4}. 
We consider as an example a global disk model with $p=-1.5$, $q=-1.0$, 
and the finite domain $-5\le z/H_{\rm c} \le 5$, where 
$H_{\rm c}/r_0=c_{{\rm s}0}/v_{{\rm K}0}$ and $c_{{\rm s}0}/v_{{\rm K}0}=0.05$.
These parameters match those presented in \citet{2013MNRAS.435.2610N}.
We solve the problem corresponding to a single radial wavenumber 
$k_x= \pm 200\pi$, associated to a wavelength of $10^{-2}r_0$,
as follows.
We discretize the problem in terms of Chebyshev cardinal functions on the Gauss-Lobatto grid, 
and the boundary condition  $\tilde{w}_z =0$ is enforced by the ``boundary bordering'' method \citep{Boyd}.
We vary the resolution, using a maximum of 300 grid points (yielding a 1200 by 1200 matrix) 
to obtain the converged eigenvalues shown. 

The basic pattern of modes shown in Figure~\ref{fig3} 
agrees well with the approximate analysis carried out by \citet{2013MNRAS.435.2610N},
who solved a second order differential equation that results from combining
and approximate Equations~(\ref{eq_VSI_general_Pi})--(\ref{eq_VSI_general_wz}) 
\footnote{\label{foot-nelson}Note that Equation~(39) of \citet{2013MNRAS.435.2610N} contains a typo.
It should have $-\sigma^2 k^2$ as the final term according to their previous equation.}.
The important difference between these two approaches is that 
 \citet{2013MNRAS.435.2610N} make an approximation which removes full 
 compressibility from the problem. 
 We see that the fastest growing modes are relatively damped when full 
 compressibility is retained in our analysis.
In order to facilitate making a connection to their findings, 
we discuss the modes that we obtained using their terminology.

The eigenvalues have the symmetry in the complex plane 
$\sigma = \sigma_r +\mathrm{sign}(k_x)|\sigma_i|$.
The fastest growing modes are a branch of ``surface'' modes with 
degenerate eigenvalues. 
In Figure~\ref{fig4} we show eigenfunctions where all components have been 
 normalized by the complex value
 \begin{align}
 \tilde{\Pi}_f(-z_m)\left[\int_{-z_m}^{z_m}| \tilde{\Pi}_f (z)/\tilde{\Pi}_f(-z_m)|^2 dz\right]^{1/2}\, 
 \end{align}
where $z_m$ is the maximum $z$ of the domain.
Because the Fourier amplitudes given by solving the 
eigenproblem  as posed in transformed-$x$ Fourier space for 
$+k_x$ and $-k_x$ are complex conjugates,
they correspond to a single real-valued eigenfunction when the $x$-direction 
Fourier transformation is inverted to bring the eigenfunction into real space.
Therefore, we have plotted this real-valued eigenfunction resulting from 
the $+k_x$ and $-k_x$ pair.
The first and second row in Figure~\ref{fig4} 
illustrate the fastest growing pair of modes with degenerate eigenvalue.
A branch of eigenvalues proceeding from the origin of the complex plane contains the 
fundamental body modes, which are associated with ``corrugation'' and ``breathing'' modes.
Like in \citet{2013MNRAS.435.2610N}, the eigenvalue closest to the origin is associated 
with the fundamental corrugation mode, shown in the third row of Figure~\ref{fig4}.
Next on this branch is the fundamental breathing mode,
 shown in the fourth row of Figure~\ref{fig4}.

\begin{figure*}
\includegraphics[width=\textwidth]{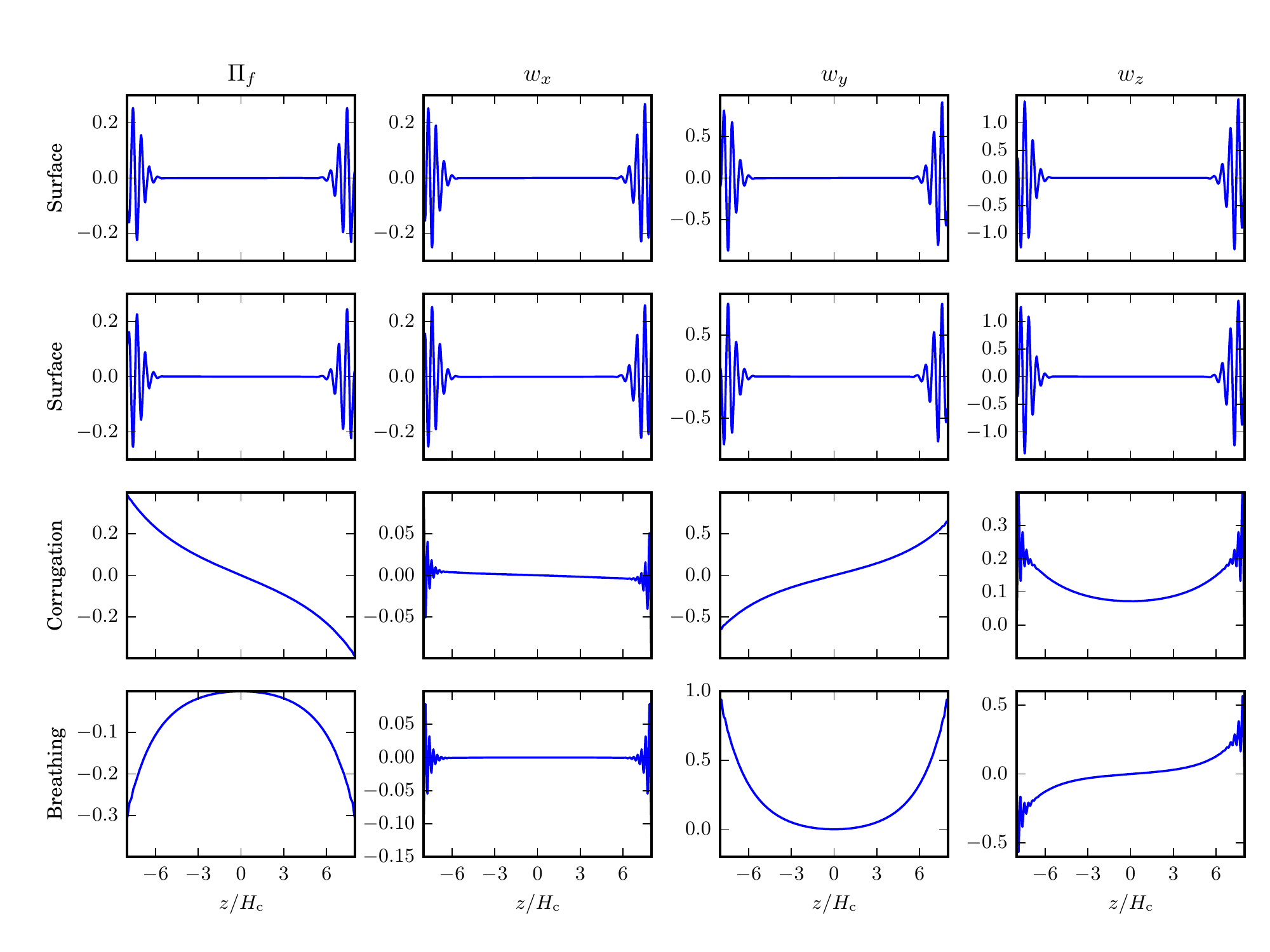}
\caption{Four example eigenfunctions, transformed to real space, of the VSI, with normalization as given in text. 
{\em Top Two Rows:} the two degenerate fastest growing surface modes with 
degenerate eigenvalues, both modes have
$\sigma \approx 0.184 \pm 0.0316 i$. 
{\em Third Row:} fundamental corrugation mode with
$\sigma \approx 0.0265 \pm 0.0307 i$.
{\em Fourth Row:} fundamental breathing mode with
$\sigma \approx 0.0376 \pm 0.0433 i$. 
}
\label{fig5}
\end{figure*}

Repeating the same calculation with a larger finite domain $-8\le z/H_{\rm c} \le 8$,
and  thicker disk $c_{{\rm s}0}/v_{{\rm K}0}=0.1$ yields instead the modes shown in 
Figure~\ref{fig5}. The remarkable change is the introduction if high
 frequency oscillations at high altitudes.
 The prodigious number of fast growing surface modes which appear in this calculation
  as the vertical size of the domain is increased suggests that magnetic fields will be a 
  significant consideration in the astrophysical context for vertical shear instabilities. 
  In the low density regime high above the disk midplane the gas will 
  almost certainly be ionized in many types of astrophysical disks; the presence of 
  magnetic fields ought then to have a significant impact on the dynamics of these instabilities.

\section{MRI in the VGSB}
\label{sec_mri}

\begin{figure}
\begin{center}
\includegraphics[width=0.45\textwidth]{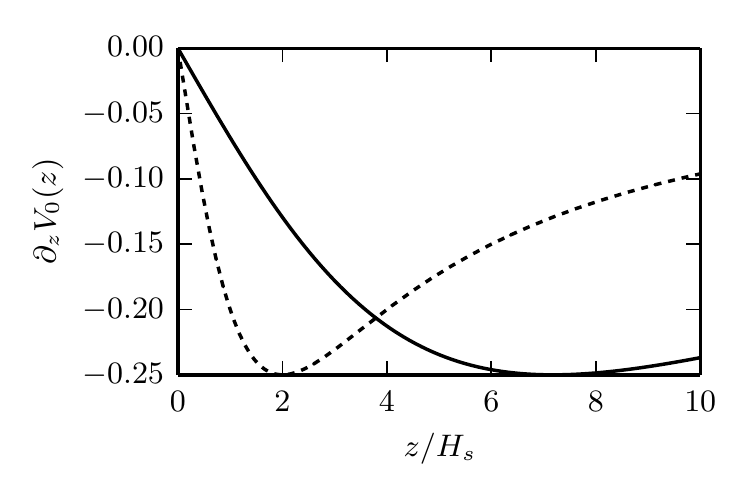}
\end{center}
\caption{Vertical shear rate in a VGSB model associated with a 
disk with a global spherical temperature structure, see Section~\ref{sec_spherical_VGSB}.
The {\sl solid line} corresponds to a thin disk with $\mu+\nu=100$, whereas 
The {\sl dashed line} corresponds to a thick disk with $\mu+\nu=10$.
}
\label{fig_mrieigen2_verticalshear}
\end{figure}

Following, and extending, the general technique of \citet{2010MNRAS.406..848L}, 
we perform a linear stability analysis of the MRI in the framework of the VGSB.
In order to derive the equations for the linear mode analysis that lead to the MRI,
we assume a homogeneous background magnetic field $\bm{B} = B_{0} \hat{\bm{z}}$
and examine the evolution of the perturbations of the form 
$f(z)\exp(\sigma t)$ for the velocity and magnetic field components $w_x$, $w_y$, $B_x$, $B_y$, 
considering $\delta \rho = w_z = 0$. Upon substitution in 
Equations~(\ref{eq_vgsb_continuity})--(\ref{eq_vgsb_energy}) and 
linearizing in the perturbation amplitudes, the equations for the perturbation are
\begin{align}
 \sigma w_x  &=  2 \Omega_0(z)  w_y  + \frac{B_0}{\mu_0\rho_0(z)}  \frac{dB_x}{dz}   \,,  
 \label{eq_MRI_VGSB_wx}
\\
 \sigma w_y  &= -\left[2\Omega_0(z) + S_0(z) \right] w_x  +
 \frac{B_0}{\mu_0\rho_0(z)} 
 \frac{dB_y}{dz} \,,  \\
\sigma B_x &= B_0 \frac{dw_x}{d z}\,,  \\
\sigma B_y &= S_0(z) B_x  + B_0 \frac{dw_y}{dz}\,.
\label{eq_MRI_VGSB_By}
\end{align}

In a similar way as we did for the VSI, it is convenient to use
the scales of length and time  provided by $r_0$ and $\Omega_0(0)^{-1}$ 
in order to define dimensionless variables, i.e., 
$\bm{x}/r_0 \rightarrow \bm{x}$ and $t\Omega_0(0)\rightarrow t$, etc.
Note that the time unit of the growth rate $\sigma$ is $1/\Omega_0(0)$, 
so its actual value will depend on the parameters of the shearing box 
model employed. 

The set of equations~(\ref{eq_MRI_VGSB_wx})--(\ref{eq_MRI_VGSB_By})
generalize Equations~(12)--(15) in \citet[][]{2010MNRAS.406..848L}
\footnote{Note a typo in  Equation~(15) of \citet[][]{2010MNRAS.406..848L}, 
where the second term on the left should include a factor of $\Omega$.}.
to include the effects of height-dependent angular frequency and shear
rate that arise in the context of baroclinic equilibrium disk models. 
For the sake of simplicity, and in order to make a direct connection
with the linear mode analysis in \citet[][]{2010MNRAS.406..848L},
we have neglected the term proportional to $B_0 \partial_z V_0$
on the right hand side of Equation (\ref{eq_MRI_VGSB_By}), which 
contributes to the secular evolution of the magnetic field. 
This is a sensible approximation if the associated modes
are localized in height in regions where the vertical shear rate $\partial_z V_0$
is smaller than the growth rate of the unstable modes.
Figure~\ref{fig_mrieigen2_verticalshear} shows the vertical shear 
rate $\partial_z V_0$
for two spherical temperature structure  VGSB models,
and the case which we consider below is roughly compatible 
with this approximation.
However, though for the thin disk case the instability growth is 
roughly ten times faster than than the vertical shear rate at $H_s=1$,
this separation is such that the possible effects warrant further 
exploration in the future. 

In order to illustrate how the MRI modes are affected, we examine 
the spectrum of growing modes present in a disk with spherical 
temperature structure, $T(\sqrt{r^2+z^2})$, as discussed in 
Section~\ref{sec_spherical_VGSB}, taking as an example the $\mu+\nu=100$.
These values corresponding to a scale height of $H_{\rm s}/r_0=\sqrt{2} c_{\rm s}(0)/v_{{\rm K}0} = \sqrt{2}$, 
and magnetic field strength 
$B_{0}/ (r_0\Omega_0(0) \sqrt{\mu_0 \rho_0(0)}\, )=0.0469$.

\begin{figure}
\begin{center}
\includegraphics[width=8cm]{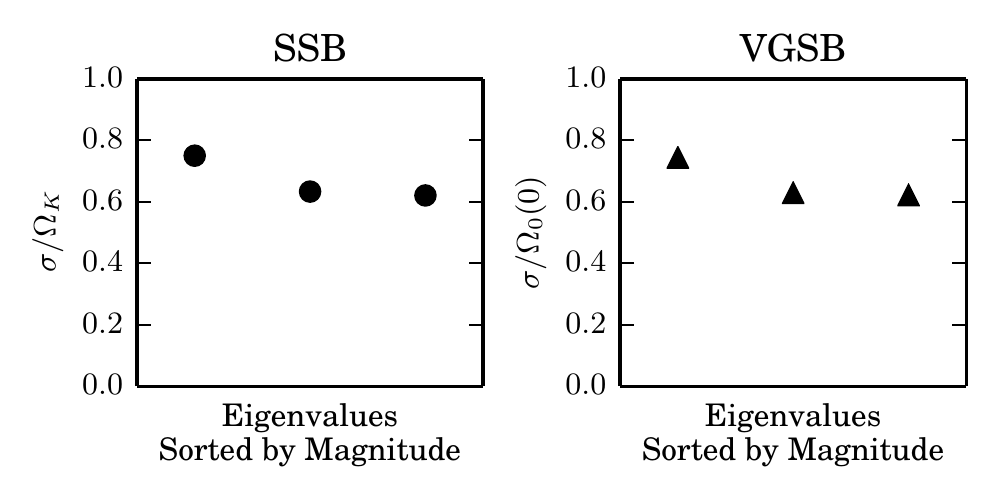}\\
\includegraphics[width=8cm]{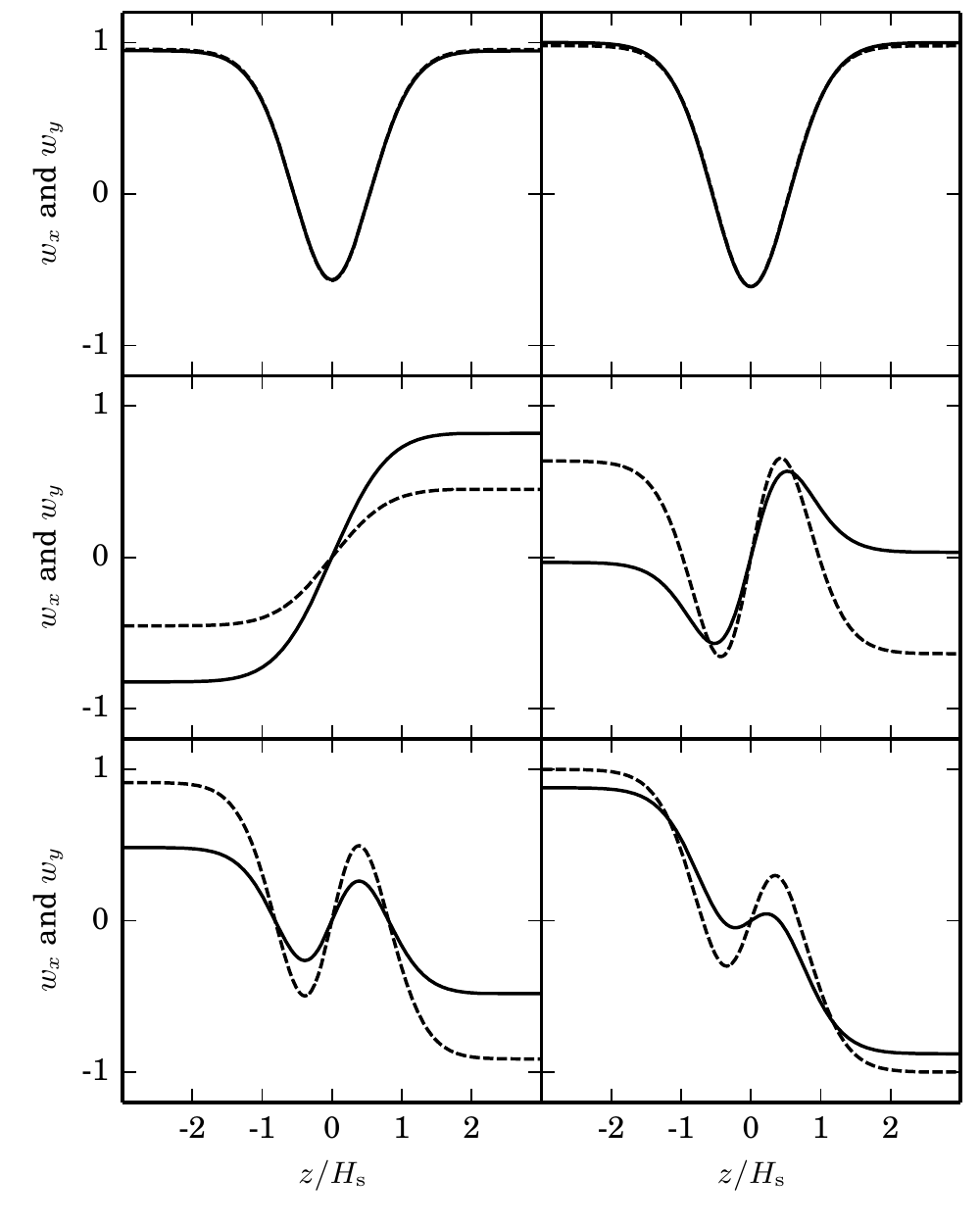}
\end{center}
\caption{Illustration of the velocity field components $w_x$ (solid line) and $w_y$ (dashed line)
corresponding to the (normalized) MRI modes, which have been ordered from top to bottom 
from fastest to slowest growth rate, as shown in the uppermost set of panels.
The results obtained in the standard shearing box used in \citet{2010MNRAS.406..848L} are shown
on the left, whereas the modes computed in the VGSB, representing a
radially local region of a disk with spherical temperature structure, are shown on the right.
}
\label{fig_mrieigen2_deltav}
\end{figure}

\begin{figure}
\begin{center}
\includegraphics[width=8cm]{fig7a}\\
\includegraphics[width=8cm]{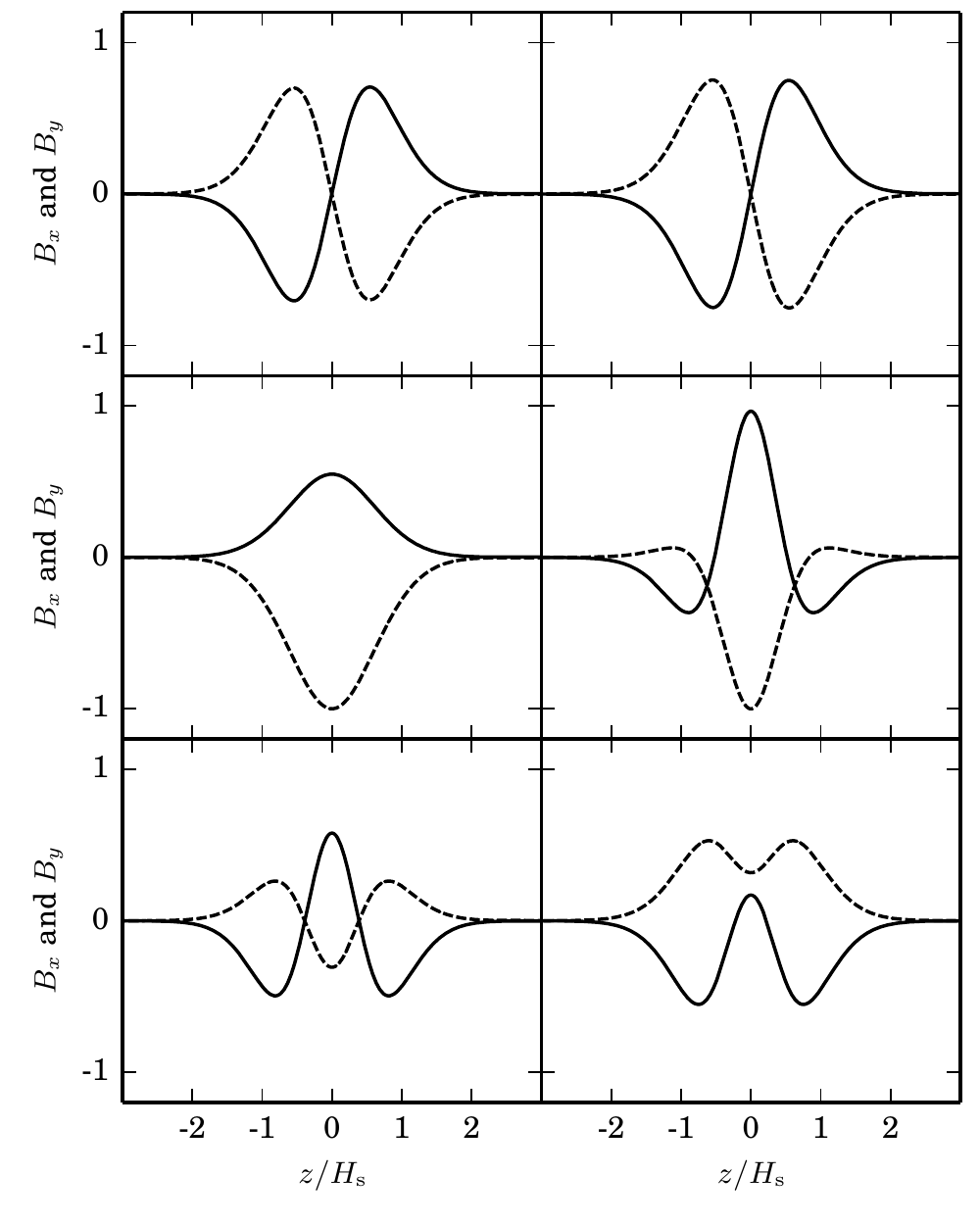}
\end{center}
\caption{
Illustration of the magnetic field components $B_x$ (solid line) and $B_y$ (dashed line)
corresponding to the (normalized) MRI modes. Other details are identical to those described
in the caption of Figure~\ref{fig_mrieigen2_deltav}.
}
\label{fig_mrieigen2_deltaB}
\end{figure}

We solve the eigenproblem posed by Equations~(\ref{eq_MRI_VGSB_wx})--(\ref{eq_MRI_VGSB_By})
by discretizing in the infinite domain $-\infty<z<\infty$ in terms of rational Chebyshev 
cardinal functions on the ``roots'', or interior grid. 
For the set of chosen parameters, 200 grid points were sufficient to obtain converged results.
We adopted the same set of 
boundary conditions used in \citet[][]{2010MNRAS.406..848L}, i.e., $B_x=B_y$ and $dw_x/dz=dw_y/dz=0$. 
To make the problem numerically tractable, we limit $\rho_0(z)$ to a minimum value of $10^{-8}$, 
which is equivalent to limiting the Alfv\'en speed at high altitudes.
The VGSB model of this particular disk model has 
one more unstable eigenmode than in the SSB.
Note that our VGSB analysis yields growth rates scaled by $\Omega_0(0)$, which is
smaller than the Keplerian frequency. Thus, even though the parameters characterizing the specific VGSB 
model do not appear explicitly in the linear analysis, the physical values obtained do depend on them.
If we consider $\mu=3$, then in the VGSB model analyzed here 
 $\Omega_0(0) = \sqrt{\nu/(\nu+\mu)}\,\Omega_{{\rm K} 0} = \sqrt{97/100}\,\Omega_{{\rm K} 0}$.
This implies that the physical growth rate of the MRI is smaller in the
 VGSB model of this particular disk than in the SSB.
In addition to solving the eigenproblem for the VGSB, we solve the 
 matching problem for the SSB by using an isothermal disk with the same 
 midplane density and scale height $H_{\rm s}$ as used in the VGSB.
The  magnetic field and velocity perturbations of the eigenmodes, shown in 
Figures~\ref{fig_mrieigen2_deltav}~and~\ref{fig_mrieigen2_deltaB}, are directionally 
orthogonal in the SSB, whereas in the  VGSB this is not the case.
The eigenmodes in the  VGSB are generally more complex, 
and, as it can be seen from these figures, they do not appear to have such a simple 
ordering in terms of the number of nodes as the SSB eigenmodes do.

\section{Discussion}
\label{sec_discussion}

The framework of the  VGSB allows, for the first time, to develop models 
for astrophysical disks which are local in radius but { are global 
in height, in that no expansion is made vertically}. 
This is critical to study astrophysical disks where the equilibrium cannot 
be described by a barotropic equation of state, as these, in general, 
do not rotate on cylinders and are thus not amenable to the SSB
 framework (see Figure~\ref{fig_tubes1}).
The  VGSB naturally accounts for 
height-dependent radial and vertical flux of fluid and 
electromagnetic momentum in the azimuthal direction.
This is relevant for the dynamics of disks with baroclinic equilibrium structure, 
especially if these cannot be regarded as thin.
Physical domains with a large vertical extent allow for 
the global magnetic field threading the disk to be efficiently 
anchored into the surrounding medium. 
The coupling between stresses and shear, both radial and vertical,
can efficiently transport momentum and energy to/from the disk. 
We envision that these effects will have important consequences, 
for example, in the study of the coupled dynamics between disks 
and their coronae and winds.

Because of its inherent local character, the framework provided by the
SSB is well suited to study physical processes involving 
scales that are smaller than the characteristic disk scales.  
Global disk simulations, which are becoming ever more accessible 
\citep{2011ApJ...735..122F,2011ApJ...738...84H}, are useful for understanding 
the large-scale disk dynamics but face the challenge of resolving the
physical processes at small scales.  In order to understand how local
and global processes interact, 
it is desirable to devise a framework to bridge local and global
approaches.  Hence, there have been several initiatives to
relax the local character of the SSB.

\subsection{Previous Works Beyond the Standard Shearing Box}

The periodic ``shearing disk" annulus introduced in \citet{2003ApJ...582..869K} 
relaxes the condition of azimuthal locality by building a radially periodic annulus 
and remapping quantities across the radial
background according to an imposed power-law radial disk structure. 
This approach involves equations which have
explicit coordinate dependence in the direction where the
computational domain is shear-periodic. Therefore, the flow
properties present a jump at the shearing boundary, which 
might lead to unphysical effects.  In this context, the
formulations presented in \citet{1996ApJ...458L..45B} and
\citet{2009A&A...498..241O} complement each other in the sense that
they retain the terms accounting for finite curvature but discard
global gradients and vice-versa,
but both lead to similar issues.

There have been several attempts to make the SSB
global in height. 
Foremost, \citet{1965MNRAS.130..125G} treated the case of a polytropic gas, 
but only made local expansions in the horizontal direction.
An early example of modifying the vertical gravity in a SSB
is given in \citet{1997ASPC..121..766M}, where the full $z$-dependence of the
vertical component of the gravitational force is considered with an isothermal gas.  
Several other authors have used
similar approaches, for both galactic
\citep[e.g.][]{1999ApJ...514L..99K} and Keplerian disks
\citep[e.g.][]{2010ApJ...718.1289S}. 
These works assume that
all the global effects in the vertical direction can be accounted for
by modifying only the momentum equation in the 
$z$-direction and exclusively through the 
term
$\partial_z \Phi_0(z) \bm{\hat{z}}$.
When the gas in the global equilibrium configuration is strictly baroptropic,
it must rotate on cylinders, and the 
height-independent shear in the SSB 
is a consistent approximation { (given the caveats in Appendix~\ref{sec:nondimensional_hydro_momentum})}.
Importantly, in the case that the gas is assumed to be isothermal, 
the SSB provides a consistent 
approach if the global disk configuration, at all radii, is isothermal.
{ If another equation of state for the hydrostatic structure
is used, or if the global configuration of which the shearing box is a 
small part has a radial temperature gradient, then the VGSB provides a 
viable framework going beyond these previous approaches, with 
important modifications beyond the SSB formalism, 
see Equations~(\ref{eq_vgsb_continuity})--(\ref{eq_vgsb_induction}).}

{
\subsection{Limitations of the VGSB}
\label{sec_limitations}

The generalization of the SSB to account for the full vertical
variation of gravity and the presence of vertical shear, inherent to
global baroclinic disk models, does posses limitations. Some of these
limitations are also characteristic of the SSB, while others are inherent to
the VGSB.

It should be clear that the VGSB, as well as the SSB, framework
consists of a set of dynamical equations for the perturbations with
respect to a time-independent background representing a smaller
section of a steady global disk model. In this approach, by
construction, the fluctuations cannot modify the dynamics of the
background. This could be considered a serious limitation, however,
this kind of approach as embodied in the SSB has proven exceptionally
useful as a workhorse to understand a wide variety of phenomena in
disks.  Another limitation of the VGSB, that is shared by the SSB, is
that the local approximation in radius prevents the full consideration
curvature terms, which are inherent to the cylindrical geometry usually 
employed to model disks (see Appendix~\ref{sec:neglecting_curvature}).

The set of equations that define the SSB can be obtained by expanding 
in radius and height either the gravitational potential of a point source 
or the Keplerian background flow that ensues when 
radial pressure support can be neglected. In order to derive the equations 
defining the VGSB we have approximated the background flow itself. The body and 
frame forces which appear in this formulation relate to a 
fluid element in that flow with the given global thermodynamics, 
not the forces on a test particle.  Thus, these forces are not the 
same as an expansion of the tidal potential about a Keplerian orbit.
In the special case of a barotropic global equilibrium, the body 
forces do correspond to a conservative potential, but in general for 
baroclinic backgrounds this is not the case.
The advantage of expanding the angular frequency is that its equilibrium
profile $\Omega(r,z)$ is sensitive to the equilibrium pressure gradient, 
enabling access to rotation laws beyond strictly Keplerian, in particular
those with vertical shear.

The most important limitations of the VGSB stem from the fact that we
have opted to explore the consequences of imposing height-dependent
shearing-periodic boundary conditions (see Section~\ref{sec_compatibility}). 
In order to do this, we had to approximate Equation~(\ref{eq_wraping}) and use
Equation~(\ref{eq_wraping_approx}) instead. Whereas this seems a
reasonable approximation, it had in general undesired consequences for
some basic conservation laws such as Kelvin's Circulation Theorem
and Alfv\'{e}n's Frozen-in Theorem unless axisymmetry is
imposed. 
In order to consider height-dependent
shearing-periodic boundary conditions we also needed to understand
under what circumstances it was acceptable to neglect background
pressure gradients. 
Even though the VGSB can be considered to be
global in height because there is no need to expand the background in
the $z$-direction, the analysis we carried out  of the pressure gradients
in Appendix~\ref{sec:nondimensional_hydro_momentum}
 shows that it is
necessary to limit the vertical extent of the domain to $z \lesssim
\sqrt{x r_0}$. This is more restrictive than $z \lesssim r_0$, but
nevertheless can accommodate for many scale-heights with the added
value of being able to account for the full expression of the vertical
gravity and vertical shear within the domain. 

The derivation of the VGSB formalism has allowed us to 
shed light into several subtle issues that are usually not addressed in the
context of the SSB, see e.g., Appendices~\ref{sec:nondimensional_hydro_momentum} and 
\ref{sec:potvort}. In spite of its limitations, the SSB has proven to be a useful 
tool to learn about local disk dynamics. We believe that the VGSB will allow the 
relaxation of some of the assumptions that have been widely adopted by using the SSB.
In particular, it will enable the investigation of some dynamical aspects of astrophysical 
disks that cannot be studied with the SSB and for which full 
global modeling is too demanding.
}

\subsection{Applications of the VGSB}
\label{sec_applications}

We anticipate that the  VGSB framework, summarized in Appendix~\ref{sec_vgsbsummary},
 will benefit the modeling of a wide variety 
of phenomena in astrophysical disks. 
The advances made possible with respect to the 
SSB will depend on the nature of the physical phenomena 
under study, e.g.,:
\paragraph{ Hydrodynamic Disk Instabilities}
The VGSB framework may
 be used for the local study of
vertical shear instabilities, such as GSF
instability \citep{1967ApJ...150..571G,1968ZA.....68..317F} and generalizations
 \citep{2013MNRAS.435.2610N}.
We have demonstrated in Section~\ref{sec_vsi} that the VGSB 
captures the correct local linear behavior.
It should also be useful to analyze the propagation  of hydrodynamic waves 
in disks \citep{1993ApJ...409..360L,1995MNRAS.272..618K}.

\paragraph{Disk Convection}
Vertical convective instabilities in disks
\citep{1980MNRAS.191...37L} have been studied with the SSB as an angular momentum transport mechanism. Though
early studies \citep{1992ApJ...388..438R,1996ApJ...465..874C,1996ApJ...464..364S}
gave negative results, \citet{2010MNRAS.404L..64L} have suggested that
earlier models are under-resolved, and outward angular momentum transport is possible.
The  VGSB framework enables to study the effects that vertical shear
can have on the long-term evolution of convective motions in a way which is 
not accessible with the SSB.

\paragraph{Disk Coronae}
Previous standard stratified shearing box simulations show the buoyant rise 
of magnetic field to the upper disk layers \citep{2000ApJ...534..398M, 
2010ApJ...713...52D}, where it is thought to dissipate giving rise to a hot corona 
\citep{1979ApJ...229..318G}. If the disk's equilibrium structure is not barotropic, 
the fact that the shear, i.e., the source of free energy, decreases with height
can have important implications for the turbulent disk dynamics and energetics 
in the disk corona. This could affect the vertical disk structure
\citep{2004ApJ...605L..45T,2008A&A...490..501J,2011ApJ...733..110B} and its corona \citep{ 
2009ApJ...704L.113B}. These effects could be important for thick disks, 
such as advection dominated accretion flows (ADAFs) \citep{1977ApJ...214..840I, 
1998tbha.conf..148N, 1999MNRAS.303L...1B}.

\paragraph{Disk Winds}
Standard shearing boxes extended in height have 
been used to study the MRI
\citep{1991ApJ...376..214B} as a mechanism for 
launching disk winds \citep{2009ApJ...691L..49S,2010ApJ...718.1289S, 
2012MNRAS.423.1318O,2012A&A...548A..76M,2013ApJ...769...76B,2013A&A...552A..71F,2013A&A...550A..61L}
though only in the framework provided by modifying the gravitational
forces through the term $\partial_z \Phi_0(z) \bm{\hat{z}}$. 
We have shown in Section~\ref{sec_mri} that the linear MRI is 
modified by effects included in the VGSB for disks with baroclinic equilibria.
Because the 
wind dynamics are particularly sensitive to the forces acting 
on the fluid far away from the midplane, the VGSB 
provides a more general framework for local studies of this nature.

\paragraph{Interstellar Medium and Galactic Disks}
In galactic disks, the VGSB makes it possible to capture critical physics
for studying the nonlinear evolution of the magnetorotational, Parker, 
and magneto-Jeans instabilities
\citep{2002ApJ...581.1080K,2005ApJ...629..849P}.  The framework 
provided by the VGSB also allows
for the inclusion of fundamental height-dependent physical effects that can 
play an important role in the study of star formation, galactic dynamos,
and the structure of the interstellar medium
\citep{2005A&A...436..585D,2006ApJ...653.1266J,2008A&A...486L..35G}.

\acknowledgements
We are grateful to Tobias Heinemann, whose valuable insights helped us 
to better understand the subtleties involved in the derivation of the VGSB
and to present it in a transparent way.
We acknowledge thoughtful comments from Scott Tremaine and Eric Blackman, 
and discussions with Troels Haugb{\o}lle, {\AA}ke Nordlund, Andrew Jackson, 
Oliver Gressel, Orkan~M.~Umurhan, Gopakumar Mohandas, Henrik Latter,
 Andrea Mignone, Gianluigi Bodo and Sacha Brun.
{ We are thankful to the anonymous referee, whose tenacious reports 
led us to better appreciate how the VGSB connects to, and extends, the SSB framework, 
and to improve the original manuscript significantly.}
The research leading to these results has received funding from the 
People Programme (Marie Curie Actions) of the European Union's Seventh 
Framework Programme  (FP7/2007-2013) under REA grant agreement 327995 (CPM), 
and the European Research Council under the European Union's Seventh Framework 
Programme (FP/2007-2013) under ERC grant  agreement 306614 (MEP). MEP also 
acknowledges support from the Young Investigator Programme of the Villum Foundation.

\appendix

\section{Derivation of the momentum equation for departures from the background bulk flow}
\label{sec_velocitysplitting}

Here, we provide the algebraic details involved in obtaining the momentum equation
(\ref{eq_mom_w}) starting from Equation~(\ref{eq_mom_before_split}).

Let us first expand the left-hand side of Equation~(\ref{eq_mom_before_split}), LHS for short, 
after substituting $\V = V\!(r,z)\hat{\bm{\phi}} + \bm{w}$, we obtain
\begin{align}
\textrm{LHS} &=
\left( \frac{\partial }{\partial t} + \V\cdot\nabla \right) \V \\
&= \frac{\partial \bm{w}}{\partial t}  + \left(V\hat{\bm{\phi}} +\bm{w}\right)\cdot\nabla \left(V\hat{\bm{\phi}} + \bm{w}\right) \\
&= \frac{\partial \bm{w}}{\partial t}
   + V\hat{\bm{\phi}} \cdot\nabla \left(V\hat{\bm{\phi}}\right) 
   + V\hat{\bm{\phi}} \cdot\nabla \bm{w} \nonumber\\
  &\quad + \bm{w}\cdot\nabla\left(V\hat{\bm{\phi}}\right) 
   + \bm{w}\cdot\nabla\bm{w} 
   \,.
\end{align}
We now expand the last three terms, recalling that the differential operators involved are in cylindrical coordinates and using the definition of the bulk flow in terms
of the angular frequency $V(r,z) = r[\Omega(r,z)-\Omega_{\rm F}]$. This yields
\begin{align}
\textrm{LHS}= &
  \frac{\partial \bm{w}}{\partial t}
   -\frac{V^2}{r}\hat{\bm{r}} \nonumber\\
& +\left[ \Omega(r,z) - \Omega_{\rm F}\right]  
        \left(
        \frac{\partial w_r}{\partial \phi} \hat{\bm{r}} +
        \frac{\partial w_\phi}{\partial \phi} \hat{\bm{\phi}} +
        \frac{\partial w_z}{\partial \phi} \hat{\bm{z}} 
        \right) \nonumber\\
&      -\frac{V w_\phi}{r}\hat{\bm{r}}
     +\frac{V w_r}{r}\hat{\bm{\phi}}  
     \nonumber \\
     & 
     - \frac{w_\phi V}{r}\hat{\bm{r}} 
     + w_r \frac{\partial V}{\partial r} \hat{\bm{\phi}} 
     + w_z \frac{\partial V}{\partial z} \hat{\bm{\phi}} 
     +\bm{w}\cdot\nabla\bm{w}     \,,
     \\
=&  \frac{\partial \bm{w}}{\partial t}
  -\Omega^2(r,z)r\hat{\bm{r}} 
  +2\Omega(r,z)\Omega_{\rm F}r\hat{\bm{r}} 
  -\Omega^2_{\rm F}r\hat{\bm{r}} \nonumber\\
&  +\left[ \Omega(r,z) - \Omega_{\rm F}\right]  
      \left(
      \frac{\partial w_r}{\partial \phi} \hat{\bm{r}} +
      \frac{\partial w_\phi}{\partial \phi} \hat{\bm{\phi}} +
      \frac{\partial w_z}{\partial \phi} \hat{\bm{z}}
      \right)
  \nonumber \\
  & 
  -2\left[\Omega(r,z)-\Omega_{\rm F}\right] w_\phi\hat{\bm{r}} 
  + \left[\Omega(r,z)-\Omega_{\rm F}\right] w_r\hat{\bm{\phi}} \nonumber\\
&  + w_r \frac{\partial r\left[\Omega(r,z)-\Omega_{\rm F}\right]}{\partial r} \hat{\bm{\phi}} 
  + w_z \frac{\partial V}{\partial z} \hat{\bm{\phi}} 
  + \bm{w}\cdot\nabla\bm{w}     \,,
   \\
=& \frac{\partial \bm{w}}{\partial t}  
  -\Omega^2(r,z)r \hat{\bm{r}}
  +2\Omega(r,z)\Omega_{\rm F}r \hat{\bm{r}} 
  -\Omega^2_{\rm F}r \hat{\bm{r}} \nonumber\\
&  +\left[ \Omega(r,z) - \Omega_{\rm F}\right]   
     \left(
     \frac{\partial w_r}{\partial \phi} \hat{\bm{r}} +
     \frac{\partial w_\phi}{\partial \phi} \hat{\bm{\phi}} +
     \frac{\partial w_z}{\partial \phi} \hat{\bm{z}}
     \right)
  \nonumber \\
& 
 +2\bm{\Omega}(r,z)\times \bm{w} 
 -2\bm{\Omega}_{\rm F} \times \bm{w}
 +w_r r \frac{\partial \Omega(r,z) }{\partial r} \hat{\bm{\phi}} \nonumber\\
& + w_z \frac{\partial r \Omega(r,z)}{\partial z} \hat{\bm{\phi}} 
 +\bm{w}\cdot\nabla\bm{w}   \,.
     \label{eq_app_lhs}
\end{align}
This completes the expansion of the left-hand side in terms of the new velocity variable $\bm{w}$.

We now proceed similarly with the right-hand side of the momentum equation~(\ref{eq_mom_before_split}), RHS for short, by substituting $\V = V\!(r,z)\hat{\bm{\phi}}+\bm{w} $ and expanding
\begin{align}
\textrm{RHS} =& 
  -\Omega^2(r,z)r\hat{\bm{r}} +\frac{1}{\rho_h}\frac{\partial P_h}{\partial r}\hat{\bm{r}} 
  +\frac{1}{\rho_h}\frac{\partial P_h}{\partial z}\hat{\bm{z}}  
  +\Omega_{\rm F}^2 r\hat{\bm{r}} \nonumber\\
&  -2\bm{\Omega}_{\rm F}\times\V  
  - \frac{\nabla P}{\rho} +\frac{1}{\rho} \bm{J}\times \bm{B} \,, \\
 =& 
   -\Omega^2(r,z)r\hat{\bm{r}} 
   +\frac{1}{\rho_h}\frac{\partial P_h}{\partial r}\hat{\bm{r}} 
   +\frac{1}{\rho_h}\frac{\partial P_h}{\partial z}\hat{\bm{z}}  
   +\Omega_{\rm F}^2 r\hat{\bm{r}} \nonumber\\
&   -2\bm{\Omega}_{\rm F}\times\left(V\hat{\bm{\phi}} +\bm{w}\right)  
   - \frac{\nabla P}{\rho} 
   +\frac{1}{\rho} \bm{J}\times \bm{B}   \,,
   \\
 =&  
  -\Omega^2(r,z)r\hat{\bm{r}} 
  +\frac{1}{\rho_h}\frac{\partial P_h}{\partial r}\hat{\bm{r}} 
  +\frac{1}{\rho_h}\frac{\partial P_h}{\partial z}\hat{\bm{z}}  
  -\Omega_{\rm F}^2 r\hat{\bm{r}} \nonumber\\
&  +2\Omega_{\rm F}\Omega(r,z)  r\hat{\bm{r}}  
  -2\bm{\Omega}_{\rm F}\times\bm{w} 
  - \frac{\nabla P}{\rho} 
  +\frac{1}{\rho} \bm{J}\times \bm{B}  
  \label{eq_app_rhs} \, .
\end{align}
Equating the left-hand side expression (Equation~\ref{eq_app_lhs}) and the right-hand side expression (Equation~\ref{eq_app_rhs}) 
and canceling out the term  $-2\bm{\Omega}_{\rm F}\times\bm{w}$ which appears on both sides,
the momentum Equation~(\ref{eq_mom_before_split}) becomes, without approximation,
\begin{align} 
\frac{\partial \bm{w}}{\partial t}  & +\left[ \Omega(r,z) - \Omega_{\rm F}\right]  
\left(
\frac{\partial w_r}{\partial \phi} \hat{\bm{r}} +
\frac{\partial w_\phi}{\partial \phi} \hat{\bm{\phi}} +
\frac{\partial w_z}{\partial \phi} \hat{\bm{z}}
\right) \nonumber\\
 &    + \bm{w}\cdot\nabla\bm{w}   + w_r r \frac{\partial \Omega(r,z) }{\partial r} \hat{\bm{\phi}} \nonumber\\
&     + w_z \frac{\partial r \Omega(r,z)}{\partial z} \hat{\bm{\phi}} 
      +2\bm{\Omega}(r,z)\times \bm{w}  \nonumber \\
      & =   \frac{\nabla P_h}{\rho_h} - \frac{\nabla P}{\rho} +\frac{1}{\rho} \bm{J}\times \bm{B}  \,,
\end{align}
which corresponds to Equation~(\ref{eq_mom_w}).

As stated in Section~\ref{sec_subtracting_bulk}, a cancellation of terms from the left and right-hand 
sides of the momentum equation results in the term 
$\bm{\Omega}(r,z)\times\bm{w}$
that looks similar 
to the Coriolis acceleration, i.e., 
$\bm{\Omega}_{\rm F}\times\bm{w}$, but with the angular frequency $\bm{\Omega}(r,z)$ 
taking the place of the constant angular frequency of the rotating frame $\bm{\Omega}_{\rm F}$.

\section{Summary of the VGSB Equations}
\label{sec_vgsbsummary}

Here we provide a self-contained summary of the equations defining the VGSB.

\paragraph{VGSB Equations}
The continuity, momentum, induction, and energy equations in the VGSB are, respectively, given by
\begin{align}
\mathcal{D}_0 \rho & +  \nabla  \cdot \left( \rho \bm{w} \right) =  0 \,, \\
\left(\mathcal{D}_0 + \bm{w} \cdot \nabla\right) \bm{w}   &+ w_z\frac{\partial V_0(z)}{\partial z}\bm{\hat{y}}
= 
  -2\Omega_0(z)\bm{\hat{z}}\times\bm{w} \nonumber\\
  -S_0(z) w_x \bm{\hat{y}} &
 - \frac{\nabla P}{\rho} -  \frac{\partial \Phi_0(z)}{\partial z}  \bm{\hat{z}}  + \frac{1}{\rho}\bm{J}\times \bm{B} \,, \\
\left(\mathcal{D}_0 + \bm{w} \cdot \nabla\right) \bm{B} &  -B_z \frac{\partial V_0(z)}{\partial z}  \bm{\hat{y}}   =
S_0(z) B_x \bm{\hat{y}} \nonumber\\
&+\left(\bm{B}\cdot\nabla\right)\bm{w} -\bm{B}\left(\nabla \cdot\bm{w}\right) \,, \label{eq_appinduc} \\
\mathcal{D}_0 e &  +  \nabla  \cdot \left( e \bm{w} \right)  = - P \left(\nabla \cdot \bm{w}\right) \,.
\end{align}
Here, $\rho$ is the mass density, $\bm{w} \equiv \bm{v} - \left[V_0(z) +S_0(z)x \right]\hat{\bm{y}} $  
is the velocity with respect to the  local bulk flow in the disk, ${\bm B}$ is the 
magnetic field, and $e$ is the internal energy density.
The pressure $P$ is determined as a function of $\rho$ and $e$, through an appropriate equation of state.
The current density is $\bm{J}\equiv \nabla\times \bm{B}/ \mu_0$,
with $\mu_0$ a constant dependent on the unit system adopted. 
All the operators in the VGSB equations are defined in a cartesian coordinate system centered at 
the fiducial radius $r_0$.

For barotropic background equilibrium disc structures, the above equations can be used in fully three dimensions,
while for baroclinic background equilibrium disc structures, these equations must by restricted to axisymmetric solutions ($y$-invariant).

\paragraph{VGSB Definitions}
The equations defining the VGSB framework depend on a number of functions of the coordinate $z$,
which result from local radial expansions of the global disk model around the radius $r_0$. 
The height-dependent advection by the background shear, $\mathcal{D}_0(z)$, is given by
\begin{align}
\mathcal{D}_0(z) \equiv \frac{\partial}{\partial t} +  \left[V_0(z) + S_0(z) x \right] \frac{\partial}{\partial y} \ ,
\end{align}
where $\Omega_0(z)$, $V_0(z)$, and $S_0(z)$ correspond, respectively, to the local values of the 
background azimuthal velocity, angular frequency, and background shear flow, which are all
derived from the angular frequency of the global disk model $\Omega(r,z)$, i.e., 
\begin{align}
V_0(z) &\equiv r_0 [\Omega_0(z)-\Omega_0(0)] \,, \\
\Omega_0(z) &\equiv \Omega(r_0,z) \,, \\
S_0(z) &\equiv r_0 \left. \frac{\partial \Omega(r,z)}{\partial r}\right|_{r=r_0} \, .
\end{align}
The gravitational potential in the VGSB $\Phi_0(z)$ is given by the value of the global
gravitational potential $\Phi(r,z)$ at $r_0$
\begin{align}
 \Phi_0(z) \equiv \Phi(r_0,z) \,.
\end{align}

\paragraph{ VGSB Boundary Conditions}
For a domain with size $L_x\times L_y \times L_z$ the horizontal boundary conditions are
\begin{align}
f(x,y,z,t) &= f(x+L_x, y + S_0(z) L_x t, z,t) \,,\\
f(x,y,z,t) &= f(x, y+L_y, z,t) \,.
\end{align}
which correspond to a generalization of the shearing-periodic boundary conditions adopted in the
SSB.

\section{Neglecting Curvature Terms}
\label{sec:neglecting_curvature}

The cylindrical coordinate system that we adopt to describe the global disk model
leads naturally to the presence of the quadratic terms 
$-w_y^2/r_0 \bm{\hat{x}}$, $w_x w_y/r_0 \bm{\hat{y}}$,
$-B_y^2/r_0 \bm{\hat{x}}$, and $B_x B_y/r_0\bm{\hat{y}}$ 
in the momentum equation (\ref{eq_mom_full}). All of these
terms are usually neglected in the SSB.
The terms related to the velocity field can be neglected safely
when they are small compared to the corresponding components of
the acceleration $-w_y\Omega_0(z) \bm{\hat{x}}$ and $w_x \Omega_0(z) \bm{\hat{y}}$.
Both of these conditions lead to the inequality
$w_y \ll r_0 \Omega_0(z)$.
The magnetic terms  $-B_y^2/r_0 \bm{\hat{x}}$ and $+B_x B_y/r_0\bm{\hat{y}}$ 
in Equation~(\ref{eq_mom_full}) are usually absent in local studies
(see \citealt{1996ApJ...458L..45B} for an exception).
If the magnetic field is sufficiently sub-thermal everywhere in the domain, 
these terms can be neglected \citep{2005ApJ...628..879P}. 
However, they have been commonly neglected even when this is not the case.
This is perhaps because it is not obvious that retaining them will lead to 
a physically consistent problem in the framework of the shearing box.
Consider, for example, the case of a shearing box in which a strong net 
azimuthal field develops as a consequence of the local disk dynamics.
This could lead to the force arising from the term $-B_y^2/r_0 \bm{\hat{x}}$ 
to increase with time. However, this force cannot be balanced by 
an increase in the centripetal acceleration as the latter is fixed by
the choice of the bulk flow \citep[e.g.][]{1995ApJ...445..337L}.
Note that all of this statements regarding the quadratic terms
that reminisce the curvilinear nature of the
original cylindrical coordinate system
are independent of whether the SSB, or the 
vertically global version developed in this paper, is considered.

\section{Nondimensionalization and Ordering of the Hydrodynamic Momentum Equation}
\label{sec:nondimensional_hydro_momentum}

A more rigorous description of the approximations that lead to the equations for the  VGSB 
can be made by introducing  nondimensional variables in order to expose the hierarchy of the 
various terms involved. 
Here, we  only deal with the hydrodynamic part of the momentum equation, for the magnetic components 
we make the assumption that the magnetic field are weak enough to not effect the global equilibrium 
background configuration. Our approach follows and  generalizes \citet{2004A&A...427..855U}.

The hydrodynamic part of the momentum equation~(\ref{momentum_w}) is given by
\begin{align}
&\left(\mathcal{D}+ \bm{w} \cdot \nabla\right) \bm{w} = -2\Omega(r,z)\hat{\bm{z}}\times\bm{w} - S(r,z) w_r \bm{\hat{\phi}} \nonumber\\
&\quad - w_z\frac{\partial V(r,z)}{\partial z}\bm{\hat{\phi}} +  \frac{\nabla P(\rho_h, e_h)}{\rho_h}   
- \frac{\nabla P(\rho, e)}{\rho} 
\label{hydro_momentum_w} \,.
\end{align}
We define a horizontal length scale $\lambda_r$, a vertical length scale $\lambda_z$,  time scale $\Omega_0^{-1}$, pressure scale 
$P_0$, and density scale $\rho_0$. Horizontal velocities are nondimensionalized by $\lambda_r \Omega_0$ and 
vertical velocities by $\lambda_z \Omega_0$. We use primes to denote normalized quantities, like $w'_r$ the radial component of 
the fluctuation velocity, and $\partial_{r'}$ the radial partial derivative with respect to the radial nondimensional length, with 
radial length nondimensionalized as $x'=x/\lambda_r$.
The components of the momentum equation are thus
 \begin{align}
&\left(\mathcal{D}'+ \bm{w}' \cdot \nabla\right) w_r' = 2\Omega'(r,z) w_\phi'  \nonumber\\
 &\qquad + \left(\frac{\epsilon}{\delta_x}\right)^2\left[ \frac{\partial_{r'}  P'(\rho_h,e_h)}{\rho'_h} - \frac{\partial_{r'}  P'(\rho,e)}{\rho'} \right] \,, \\
&\left(\mathcal{D}'+ \bm{w}' \cdot \nabla\right) w'_\phi = -2\Omega'(r,z) w_r'  - S'(r,z) w'_r \nonumber\\
  &\qquad + \left(\frac{\epsilon}{\delta_x}\right)^2\left[  - \frac{(1/r')\partial_{\phi} P'(\rho,e)}{\rho'} \right]  \,,  \\
&\left(\mathcal{D}'+ \bm{w}' \cdot \nabla\right) w'_z = \nonumber\\
&\qquad   \left(\frac{\epsilon}{\delta_z}\right)^2\left[\frac{\partial_{z'} P'(\rho_h,e_h)}{\rho'_h} - \frac{\partial_{z'} P'(\rho,e)}{\rho'} \right]    \,.
\label{nondim_hydro_momentum_wx} 
\end{align}
where we have introduced the dimensionless parameters
\begin{align}
\epsilon &\equiv \frac{1}{\Omega_0 r_0} \sqrt{ \frac{P_0}{\rho_0} } \,, \\
\delta_x &\equiv \frac{\lambda_r}{r_0} \,, \\
\delta_z &\equiv \frac{\lambda_z}{r_0}  \,.
\end{align}

Note that $\Omega_0$ is most readily identified as the $\Omega(r,z)$ appearing at the location 
of the dynamics in question, not the Keplerian frequency. 
Thus $\epsilon$ is not directly the usual nondimensional scale height.
Taylor expanding in small $\delta_x$ about $r_0$, keeping only lowest order terms, 
and transforming to cartesian coordinates as this allows 
(Section~\ref{sec_approx_in_horiz_planes}) yields
 \begin{align}
&\left(\mathcal{D}'_0+ \bm{w}' \cdot \nabla\right) w_x' = 2\Omega'_0(z) w_y' \nonumber\\
&\qquad  + \left(\frac{\epsilon}{\delta_x}\right)^2\left[\frac{\partial_{x'} P'(\rho_h,e_h)}{\rho'_h} - \frac{\partial_{x'} P'(\rho,e)}{\rho'} \right] \,,\\
&\left(\mathcal{D}'_0+ \bm{w}' \cdot \nabla\right) w'_y = -2\Omega'_0(z) w'_x  - S'_0(z) w'_x \nonumber\\
&\qquad   + \left(\frac{\epsilon}{\delta_x}\right)^2\left[ - \frac{\partial_{y'} P'(\rho,e)}{\rho'} \right]  \,, \\
&\left(\mathcal{D}'_0+ \bm{w}' \cdot \nabla\right) w'_z = \nonumber\\
&\qquad    \left(\frac{\epsilon}{\delta_z}\right)^2\left[ \frac{\partial_{z'} P'(\rho_h,e_h)}{\rho'_h} - \frac{\partial_{z'} P'(\rho,e)}{\rho'} \right] \,.
\end{align}
We now decompose the pressure and density into their hydrostatic part and a their associated fluctuation as
\begin{align}
P' &\equiv P_1'+P'_h \,, \\
\rho' &\equiv \rho'_1 + \rho'_h \,.
\end{align}
The components of the momentum equation then become
\begin{align}
&\left(\mathcal{D}'_0+ \bm{w}' \cdot \nabla\right) w_x' = 2\Omega'_0(z) w_y' \nonumber\\
 &\qquad + \left(\frac{\epsilon}{\delta_x}\right)^2\left[   \frac{ \rho'_1 \partial_{x'} P'_h}{\rho'_h ( \rho'_h + \rho'_1 )} - \frac{ \partial_{x'} P'_1}{ \rho'_h + \rho'_1 }  \right] \,,\\
&\left(\mathcal{D}'_0+ \bm{w}' \cdot \nabla\right) w'_y = -2\Omega'_0(z) w'_x  - S'_0(z) w'_x \nonumber\\
 &\qquad + \left(\frac{\epsilon}{\delta_x}\right)^2\left[ - \frac{ \partial_{y'} P'_1}{ \rho'_h + \rho'_1 } \right]   \,, \\
&\left(\mathcal{D}'_0+ \bm{w}' \cdot \nabla\right) w'_z = \nonumber\\
 &\qquad   \left(\frac{\epsilon}{\delta_z}\right)^2\left[  \frac{ \rho'_1 \partial_{z'} P'_h}{\rho'_h ( \rho'_h + \rho'_1 )} - \frac{ \partial_{z'} P'_1}{ \rho'_h + \rho'_1 } \right] \,.
\end{align}
The fluctuation pressure terms have the same scaling as the background hydrostatic pressure terms, 
as in \citet{2004A&A...427..855U}.   Using the expression for hydrostatic equilibrium given by Equation~(\ref{eq_omegarz}), 
we obtain the form of the radial and vertical components of the acceleration as
\begin{align}
&\left(\frac{\epsilon}{\delta_x}\right)^2 \frac{1}{\rho_h'} \frac{\partial P_h'}{\partial x'}  =\nonumber\\
&\quad   -\frac{(1+\delta_x x')}{ \delta_x} \left[\Omega'^2 - \Omega_{{\rm K}0}'^2 \left[ (1+\delta_x x')^2 +\delta_z^2 z'^2\right]^{-3/2}\right] \,,
\end{align}
\begin{align}
 \left(\frac{\epsilon}{\delta_z}\right)^2\frac{1}{\rho'_h} \frac{\partial P'_h}{\partial z'} &=  
     - \Omega_{{\rm K}0}'^2 z' \left[ (1+\delta_x x')^2 + (\delta_z z')^2\right]^{-3/2} \,.
\end{align}
The vertical acceleration depends only on the gravitational potential, but the radial acceleration
depends on the thermodynamics of the hydrostatic equilibrium though the nondimensionalized rotation $\Omega'$.

In order to illustrate the procedure for determining the appropriate expansion of the radial acceleration, we work with the structure 
of the hydrostatic background provided by the cylindrical  temperature profile $T(r)$, as in 
Section~\ref{sec_cylindrical_VGSB}. This thermal structure contains a barotropic equilibrium as a special case.
The nondimensional version of the rotation law given by Equation~(\ref{omega_T_cyl}) is
\begin{align}
\Omega'^2 =& \Omega_{{\rm K}0}'^2 (1+\delta_x x')^{-3} \Bigg(+ (p+q)\epsilon^2(1+\delta_x x')^{q+1}  \nonumber\\
     &+q \left[1-\frac{(1+\delta_x x')}{\sqrt{(1+\delta_x x')^2 +\delta_z^2 z'^2}}  \right]  \Bigg) \,.
\end{align}
When $q=0$ this structure is isothermal, and barotropic, and when $\epsilon \ll 1$ the disk is thin, with the rotation 
close to Keplerian.

{ \subsection{Case of SSB and  VGSB for a Globally Isothermal Background}}

In order to connect our analysis with \citet{2004A&A...427..855U}, who considered the case with $\epsilon=0$ and 
$\delta_x=\delta_z$, we examine explicitly here the globally isothermal case with $q=0$.
Even though the equations that we presented for the VGSB do not explicitly involve expansions 
in the vertical direction, at this point in the analysis we do make such an expansion in order to assess the 
conditions under which it is acceptable to neglect the radial pressure gradients that need to be discarded 
in order to impose shearing-periodic boundary conditions, as discussed in Section \ref{sec_compatibility}.
{ Thus, the results of this section apply equally to the VGSB model for a globally isothermal (barotropic) 
disk, as well as to the SSB.} 

Expanding the radial and vertical components of the acceleration about $\delta_x=0$ and $\delta_z=0$  we
obtain, to leading order, 
\begin{align}
\Omega_{{\rm K}0}'^{-2}  \left(\frac{\epsilon}{\delta_x}\right)^2 \frac{1}{\rho_h'} \frac{\partial P_h'}{\partial x'}  \simeq 
   -\frac{p\epsilon^2}{\delta_x}  +p \epsilon^2 x' - p\epsilon^2  x'^2\delta_x \nonumber\\
    + p \epsilon^2 x'^3 \delta_x^2 
     - \frac{3 z'^2}{2 \delta_x}\delta_z^2 +6 x' z'^2 \delta_z^2 + \ldots \,, \label{eq_radaccbaro}
\end{align}
\begin{align}
\Omega_{{\rm K}0}'^{-2}  \left(\frac{\epsilon}{\delta_z}\right)^2 \frac{1}{\rho_h'} \frac{\partial P_h'}{\partial z'}  \simeq 
  -z' + 3x'z'\delta_x + \frac{3}{2}z'^3 \delta_z^2 + \ldots \,. \label{eq_vertaccbaro}
\end{align}
It is evident, { as any finite domain contains arbitrarily small $\delta_x$,} that the only rigorous limit in which we can strictly justify discarding every term on the right hand 
side of equation (\ref{eq_radaccbaro}) is the one corresponding to a disk with a constant midplane density, $p=0$, and/or 
infinitely thin, $\epsilon=0$, in which the characteristic scales of interest in the vertical direction vanish 
identically, $\delta_z=0$. This is certainly not an interesting limit, especially in the case where $\delta_x=\delta_z$.
{ Perhaps this is} the reason for which these terms are usually discarded in the standard shearing box formalism, even 
though they cannot be strictly neglected in a disk with a radial density gradient and/or non-vanishing scale height 
when the vertical extent of the domain does not vanish identically.

Having this caveat in mind, let us examine under what conditions we can argue that the terms proportional to 
$\delta_x$ and $\delta_z$ are sufficiently small that it is acceptable to neglect them. This requires understanding 
when each of the dimensionless terms in question is small compared to unity.
In equation (\ref{eq_radaccbaro}) for the radial acceleration, the first term can be argued to be sufficiently 
small for $\delta_x  \ll 1$ provided that $p \epsilon^2 / \delta_x  \ll 1$, which implies 
$p \ll \rho_0 (r_0^2\Omega_0^2)/P_0 \sim (r_0\Omega_0/c_{\rm s})^2$. 
This condition can be satisfied for power-law indices $p$ of order unity.
{ The second term, of order $\delta_x^0$, is small based on the same condition as $x'$ is at most of order unity.}
The third and fourth terms are clearly small for $\delta_x  \ll 1$.
The { fifth} term in this equation is more difficult to deal with because it scales with the ratio $\delta_z^2/\delta_x$. 
In the case considered by  \citet{2004A&A...427..855U}, because $\delta_x=\delta_z$, 
this term is simply proportional to $\delta_x$, and it can be directly neglected in 
the limit $\delta_x \ll 1$, along with the higher order terms in the radial acceleration.
When the assumption $\delta_x=\delta_z$ is relaxed, the second term in the radial acceleration is small when 
$z'^2  \delta_z^2 \ll \delta_x$. Restoring dimensional quantities, this condition becomes 
$z \ll \sqrt{\lambda_x r_0} <  \sqrt{x r_0}$. This implies that the fifth term 
in the equation for the radial acceleration is small provided that the vertical extent of the domain is smaller
than the geometric mean between its radial extent and the radial location of the box. 
Note that this condition is more restrictive that requesting that $z < r_0$, but far less restrictive
than imposing that $z$ vanishes identically.
The higher order terms, such as the sixth term on the right-hand side of equation (\ref{eq_radaccbaro}), 
are negligible provided that $x \ll r_0$.
The vertical acceleration in equation (\ref{eq_vertaccbaro}) does not depend on the parameters 
$p$ or $\epsilon$. In this case, it is straightforward to neglect higher order terms proportional 
to $\delta_x$ and $\delta_z$ with respect to the leading term $z'$.

{ \subsection{Case of VGSB for Cylindrical Temperature Profiles}

Here, we generalize the above result in order to include a baroclinic background with non-zero radial temperature power law index $q$.
The generalization of Equation~(\ref{eq_radaccbaro}) is
\begin{align}
&\Omega_{{\rm K}0}'^{-2} \left(\frac{\epsilon}{\delta_x}\right)^2 \frac{1}{\rho_h'} \frac{\partial P_h'}{\partial x'} \simeq 
  -\frac{(p+q) \epsilon^2 }{ \delta_x } 
  +(1-q)(p+q)\epsilon^2 x'  \nonumber\\
&\qquad- \frac{1}{2}(q-2)(q-1)(p+q)\epsilon^2 x'^2  \delta_x \nonumber\\
&\qquad -\frac{1}{6}(q-3)(q-2)(q-1)(p+q)\epsilon^2 x'^3 \delta_x^2 \nonumber\\
&\qquad -(3+q)\frac{z'^2 \delta_z^2}{2 \delta_x}  +(6+2q)x'z'^2\delta_z^2
   + \ldots \, . \label{radaccbaroclin}
\end{align}
It is clear that there is a one-to-one correspondence between each 
of the terms present in Equation~(\ref{radaccbaroclin}) and the ones
appearing in Equation~(\ref{eq_radaccbaro}). For any reasonable value of $q$, 
the numerical coefficients in both equations are of the same order.
Because of this, the conditions required to neglect each of the terms
in Equation~(\ref{radaccbaroclin}) and Equation~(\ref{eq_radaccbaro})
are, within factors of order unity, identical. The equation for the
expansion of the vertical pressure gradient is independent of $q$ and
thus identical to Equation~(\ref{eq_vertaccbaro}). 

\vspace{0.5cm}

From the analysis in this appendix, we thus conclude that, within factors of order unity, 
the requirements to neglect the radial background pressure gradients in the VGSB are 
expected to be similar to the ones involved in the SSB.}

\section{Potential Vorticity and Ertel's Theorem}
\label{sec:potvort}

In previous sections of this paper, we have addressed the impact that the approximations embodied 
in the 
VGSB framework have on Kelvin's Circulation Theorem and Alfv\'{e}n's Frozen-in Theorem and showed that 
these are satisfied if the fluid is barotropic or axisymmetric. Another important conservation law is given by Ertel's Theorem, which governs the evolution of the potential vorticity 
$(\bm{\omega}_{a}\cdot \nabla A)/{\rho}$, where $\bm{\omega}_{a} = \nabla \times\bm{v} + 2 \bm{\Omega}_{\rm F}$ is the absolute vorticity and $A$ is a fluid property advected with the flow, according to  \citep{1982bsv..book.....P}
\begin{align}
\left(\frac{\partial}{\partial t} + \bm{v} \cdot \nabla\right) \left[\frac{ \bm{\omega}_{a}}{\rho} \cdot \nabla A \right]
=   \frac{\nabla \rho \times \nabla P }{\rho^3}  \cdot \nabla A \,.
\label{eq:Ertel_full}
\end{align}
{
Note that for any conserved scalar field $A$, Ertel's Theorem leads to a conservation law for the potential vorticity 
if the flow is barotropic, i.e.,  $\nabla \rho \times\nabla P=0$.  If $A$ is taken to be the specific entropy $s$
 in an
 isentropic flow, i.e. $(\partial /\partial t+\bm{v}\cdot\nabla) s = 0$, Ertel's Theorem provides a conservation law 
for the potential vorticity even if the flow is baroclinic, because $(\nabla \rho \times \nabla P) \cdot  \nabla s =0$.

In order to shed light into the implications that the approximations embodied in the VGSB have for Ertel's Theorem
let us examine the evolution equation for a PV-type quantity defined, in terms of the VGSB background velocity field
$\bm{V}(x,z) = [V_0(z)+xS_0(z)]\bm{\hat{y}}$ and the velocity fluctuations $\bm{w}$, as 
\begin{align}
\frac{\bm{\omega}_a}{\rho}&\cdot \nabla A = \nonumber\\
&\frac{1}{\rho}\left\{\nabla\times[V_0(z)\bm{\hat{y}} + S_0(z) x \bm{\hat{y}}  + \bm{w}] + 2\bm{\Omega}_{\rm{F}}\right\} \cdot \nabla A \label{eq_full_pv_def} \ .
\end{align}
We can assess when such a quantity obeys a conservation law in the form of Equation~(\ref{eq:Ertel_full}).
Starting from the approximate VGSB momentum equation~(\ref{eq_vgsb_momentum}), transforming from the fluctuation velocity $\bm{w}$ to the velocity $\bm{v} = [V_0(z)+xS_0(z)]\bm{\hat{y}} + \bm{w}$, and following the usual 
steps for deriving a potential vorticity evolution equation \citep{1982bsv..book.....P}, we arrive to the following evolution equation
\begin{align}
\Bigg( \frac{\partial }{\partial t} &+\bm{v}\cdot \nabla\Bigg)  \left[\frac{\bm{\omega}_a}{\rho} \cdot \nabla A \right]= \nonumber\\
& \Bigg[\frac{1}{\rho}\nabla\times \bigg\{ 2\Omega_{\rm F}\bm{\hat{z}} \times \left[ V_0(z) + S_0(z)x\right] \bm{\hat{y}} 
      \nonumber\\
&
+2(\Omega_{\rm F}-\Omega_0(z))\bm{\hat{z}}\times\bm{w}
+ w_z \frac{\partial S_0(z)}{\partial z} x\bm{\hat{y}}
 \bigg\}\Bigg] \cdot \nabla A \nonumber\\
&  +\frac{\nabla\rho \times\nabla P}{\rho^3} \cdot\nabla A \label{eq_regular_pv_vgsb}\ .
\end{align}
The first term on the right-hand side can be non-zero, so this is not in general a conservation law of the form Equation~(\ref{eq:Ertel_full}).
This  term has three components within it, the first being a nonconservative tidal force, 
the second being a coriolis-like force which remains from the velocity transformation,
 and the final term being due to the approximation made in Equation~(\ref{eq_wraping_approx}).
A coriolis-like force remains from the velocity transform from $\bm{w}$ to $\bm{v}$ because in Cartesian coordinates it is just a linear velocity boost in $\bm{\hat{y}}$, and 
the VGSB momentum equation contains a coriolis-like term which is proportional to the $\bm{\hat{y}}$ velocity.
Interestingly, when a velocity transformation is done in cylindrical coordinates with an analogous change in  the $\bm{\hat{\phi}}$ velocity, 
the matching change in the coriolis term cancels with components of the cylindrical coordinate advection operator on the left-hand side, 
as was seen in Appendix~\ref{sec_velocitysplitting}.
In the case of a VGSB model with a height-independent $\Omega_0(z)$ or an SSB, the 
first term on the right-hand side of Equation~(\ref{eq_regular_pv_vgsb})  is zero, as the first two 
components are conservative and curl-free, and the third component is zero. 
In those cases, this applies for any choice of $\Omega_{\rm F}$, as $\Omega_0$ is constant.
Thus in those cases this PV-type quantity is conserved in the 
same sense as in Equation~(\ref{eq:Ertel_full}).
Beyond simply not being in general in the form of a conservation law, Equation~(\ref{eq_regular_pv_vgsb}) also has the property that the Lagrangian derivative term with $\partial /\partial t+\bm{v}\cdot\nabla$ is not zero when the velocity fluctuations $\bm{w}$ are zero. This is because the quantity defined in Equation~(\ref{eq_full_pv_def}) contains both the background velocity and the fluctuation velocity. Note that in the absence of fluctuations in an 
axisymmetric flow, i.e., $\partial_y \equiv 0$, and the first term on the right-hand side of Equation~(\ref{eq_regular_pv_vgsb}) does vanish.

Understanding the implications of the source terms that would appear in general on the right-hand side of Equation~(\ref{eq_regular_pv_vgsb}) is beyond the scope of this work. In the reminder of this section we have a more modest goal which consists of assessing how the approximations leading to the VGSB impact the dynamics of a PV-type quantity 
$(\delta \bm{\omega} \cdot \nabla A)/\rho$ defined solely in terms of the velocity fluctuations, $\bm{w}$.
}

In a  way similar to that in the SSB, the VGSB framework consists of a
set of equations for the fluctuations with respect to a known local
equilibrium background (which corresponds to a local approximation of
a global equilibrium). 
Starting from the momentum equation in terms of the fluid velocity $\bm{v}$
in the frame rotating with $\bm{\Omega}_{\rm F}$, we have derived the exact
Equation~(\ref{momentum_w}) for the velocity $\bm{w}$ characterizing the departures from the
background equilibrium $\bm{V}(r,z)  = r\left[\Omega(r,z) - \Omega_{\rm F} \right]\hat{\bm \phi}$, Equation~(\ref{eq_Vrz}). 
In what follows, we derive an evolution equation for the PV-type quantity associated with the velocity fluctuations in
an exact way (Equation~\ref{exact_ertel_w}), and then repeat the derivation for the VGSB 
approximation (Equation~\ref{vgsb_ertel_w}) with the goal of comparing both results.

Starting from the exact momentum equation~(\ref{momentum_w}), 
taking the curl and using the continuity equation, we arrive to
\begin{align}
\frac{1}{\rho}&(\mathcal{D}+ \bm{w} \cdot \nabla) \delta \bm{\omega}
    +\left(\frac{\delta \bm{\omega}}{\rho}\right) \nabla \cdot \bm{w} 
        +\frac{1}{\rho} \nabla \Omega(r,z) \times \frac{\partial\bm{w} }{\partial \phi}  
     = \nonumber\\
    & \frac{1}{\rho}\bm{w}[\nabla \cdot 2\Omega(r,z)\bm{\hat{z}}] - \frac{1}{\rho} 2\Omega(r,z) \bm{\hat{z}}(\nabla \cdot \bm{w})
      \nonumber\\
&  +\left(\frac{\delta \bm{\omega} }{\rho}   \cdot\nabla\right) \bm{w} 
    +\left(\frac{ 2\Omega(r,z) \bm{\hat{z}} }{\rho}   \cdot\nabla\right) \bm{w} \nonumber\\
& - \frac{1}{\rho}\nabla \times \left(w_z\frac{\partial V(r,z)}{\partial z}\bm{\hat{\phi}} +S(r,z) w_r \bm{\hat{\phi}} \right)   \nonumber\\
& + \frac{\nabla P(\rho_h, e_h)}{\rho} \times \nabla \frac{1}{\rho_h} - \frac{\nabla P}{\rho} \times \nabla \frac{1}{\rho} \, ,
\label{eq_mom_curl_div_rho}
\end{align}
where it has been natural to define the quantity 
$\delta \bm{\omega} \equiv \nabla\times\bm{w} $, i.e., the vorticity 
associated with the fluctuations.
We can obtain the following relation from the continuity equation~(\ref{eq_cont})
\begin{align}
\frac{ \delta \bm{\omega} }{\rho} \nabla\cdot \bm{w} 
  &=  -\frac{ \delta \bm{\omega} }{\rho^2} (\mathcal{D} + \bm{w}  \cdot \nabla) \rho
 \, ,
\end{align}
and use it to replace the second term on the left-hand side in Equation~(\ref{eq_mom_curl_div_rho}) to obtain
\begin{align}
(\mathcal{D}&+ \bm{w} \cdot \nabla)\left(\frac{\delta \bm{\omega}}{\rho} \right)
  +\frac{1}{\rho} \nabla \Omega(r,z) \times \frac{\partial  \bm{w}}{\partial \phi} 
       \nonumber\\
   & + \frac{1}{\rho}\nabla \times \left( w_z\frac{\partial V(r,z)}{\partial z}\bm{\hat{\phi}} + S(r,z) w_r \bm{\hat{\phi}} \right)
    =  \nonumber\\
&
    \left(\frac{\delta \bm{\omega}}{\rho} \cdot\nabla \right) \bm{w}
    +\left(\frac{ 2\Omega(r,z) \bm{\hat{z}} }{\rho}   \cdot\nabla\right) \bm{w} \nonumber\\
&    + \frac{\bm{w} }{\rho}[ \nabla \cdot 2\Omega(r,z)\bm{\hat{z}} ]
    - \frac{2\Omega(r,z)}{\rho} \bm{\hat{z}}(\nabla \cdot \bm{w}) \nonumber\\
 &    + \frac{\nabla P(\rho_h, e_h)}{\rho} \times \nabla \frac{1}{\rho_h} 
    - \frac{\nabla P}{\rho} \times \nabla \frac{1}{\rho}  \, .
    \label{exact_vortensity}
\end{align}
If we consider a scalar quantity $A$ which is frozen-in to the flow, i.e., 
$(\mathcal{D}+\bm{w}\cdot\nabla) A = 0$, 
we can derive the following identity
\begin{align}
\frac{\delta \bm{\omega}}{\rho} &\cdot  \nabla [ (\mathcal{D}+\bm{w}\cdot\nabla) A] = 
 \frac{\delta \bm{\omega}}{\rho} \cdot  [(\mathcal{D}+\bm{w}\cdot\nabla) \nabla A]  \nonumber\\
 &+  \frac{\delta \bm{\omega}}{\rho} \cdot  [\nabla \Omega(r,z)]\partial_\phi A 
 + \left[\left(\frac{\delta \bm{\omega}}{\rho} \cdot \nabla \right)  \bm{w}\right]\cdot  \nabla A = 0\,.
\label{exact_advection}
\end{align}
We can combine equations (\ref{exact_vortensity}) and (\ref{exact_advection}) 
to obtain an evolution equation 
for $(\delta \omega \cdot \nabla A)/\rho$
as
\begin{align}
&(\mathcal{D}+ \bm{w} \cdot \nabla)   \left[\frac{\delta \bm{\omega} }{\rho} \cdot \nabla A \right]
    +\left[ \frac{1}{\rho} \nabla \Omega(r,z) \times \frac{\partial \bm{w}}{\partial \phi} \right]\cdot \nabla A \nonumber\\
& + \left[\frac{1}{\rho}\nabla \times \left( w_z\frac{\partial V(r,z)}{\partial z}\bm{\hat{\phi}}  + S(r,z) w_r \bm{\hat{\phi}} \right) \right] \cdot \nabla A
      \nonumber\\
&= 
  \bigg[\left(\frac{ 2\Omega(r,z) \bm{\hat{z}} }{\rho} \cdot \nabla \right)  \bm{w} 
           +\frac{\bm{w}}{\rho}(\nabla \cdot 2\Omega(r,z)\bm{\hat{z}} )\nonumber\\
&           - \frac{2\Omega(r,z)}{\rho} \bm{\hat{z}}(\nabla \cdot \bm{w})  \bigg]\cdot  \nabla A 
   - \left(\frac{\delta \bm{\omega} }{\rho} \cdot  \nabla \Omega(r,z) \right) \frac{\partial A}{\partial \phi} 
       \nonumber\\
&  
  + \left[ -\frac{\nabla \rho_h \times \nabla P(\rho_h, e_h)  }{\rho \rho_h^2} 
           + \frac{ \nabla \rho \times \nabla P }{\rho^3}  \right] \cdot \nabla A  \, .
           \label{exact_ertel_w}
\end{align}This result is exact and it follows directly from Equations~(\ref{eq_cont}) and (\ref{eq_mom}) in this paper, provided that $A$ is frozen-in to the flow.

Following the same procedures in the VGSB context, starting from the approximate VGSB momentum equation~(\ref{eq_vgsb_momentum}) and again taking the curl
and combining with the continuity equation we arrive at
\begin{align}
(\mathcal{D}_0 &+ \bm{w} \cdot \nabla) \left(\frac{\delta \bm{\omega}_{0}}{\rho}\right)
   +  \frac{1}{\rho}\nabla \left[ V_0(z) + S_0(z) x  \right] \times \frac{\partial   \bm{w} }{\partial y} \nonumber\\
&   + \frac{1}{\rho}\nabla\times\left( w_z\frac{\partial V_0(z)}{\partial z}\bm{\hat{y}} + S_0(z) w_x \bm{\hat{y}}\right) \nonumber\\
= %
 &
  \left(\frac{\delta \bm{\omega}_{0}}{\rho} \cdot \nabla\right)\bm{w} 
   +\left(\frac{ 2\Omega_0(z) \bm{\hat{z}} }{\rho}   \cdot\nabla\right) \bm{w} \nonumber\\
&    +\frac{1}{\rho}\bm{w}( \nabla \cdot 2\Omega_0(z)\bm{\hat{z}} ) 
    - \frac{1}{\rho} 2\Omega_0(z) \bm{\hat{z}}(\nabla \cdot \bm{w}) \nonumber\\
&
   - \frac{1}{\rho}{\nabla P} \times \nabla \frac{1}{\rho} \, .
       \label{vgsb_vortensity}
\end{align}
Here, it has also been natural to define $\delta \bm{\omega}_{0} \equiv \nabla\times\bm{w}$,
i.e., the voticity
associated with the fluctuations with respect to the VGSB background.
This result is the VGSB equivalent of Equation~(\ref{exact_vortensity}).
The notable difference between the VGSB and the exact form is the lack of the 
  $\nabla \rho_h\times\nabla P_h$ term.
Considering a scalar quantity frozen-in to the approximate background 
$(\mathcal{D}_0+\bm{w}\cdot\nabla) A = 0$, we obtain
\begin{align}
(\mathcal{D}_0 & + \bm{w} \cdot \nabla) \left[ \frac{\delta \bm{\omega}_{0}}{\rho} \cdot\nabla A \right] \nonumber\\
&      +  \left[ \frac{1}{\rho}\nabla \left[ V_0(z) + S_0(z) x  \right] \times \frac{\partial \bm{w}}{\partial y} \right] \cdot\nabla A \nonumber\\
&     + \left[ \frac{1}{\rho}\nabla\times\left( w_z\frac{\partial V_0(z)}{\partial z}\bm{\hat{y}} +  S_0(z) w_x \bm{\hat{y}} \right) \right] \cdot\nabla A
       \nonumber\\
= &
   \Bigg[\left(\frac{ 2\Omega_0(z) \bm{\hat{z}} }{\rho} \cdot \nabla \right)  \bm{w} 
         +  \frac{\bm{w}}{\rho}( \nabla \cdot 2\Omega_0(z)\bm{\hat{z}} ) \nonumber\\
&          - \frac{2\Omega_0(z) }{\rho}  \bm{\hat{z}}(\nabla \cdot \bm{w}) 
     \Bigg]\cdot  \nabla A \nonumber\\
&   -   \left(\frac{\delta \bm{\omega}_{0}}{\rho} \cdot  \nabla [V_0(z) + xS_0(z)] \right) \frac{\partial  A}{\partial y} 
    \nonumber\\
  & 
  +\left[ 
  \frac{\nabla\rho \times \nabla P }{\rho^3} \right] \cdot\nabla A  \, .
 \label{vgsb_ertel_w}
\end{align}
This result is the VGSB version of the exact result in Equation~(\ref{exact_ertel_w}). 
From this form, we can see that in the shearing box (both SSB and VGSB) the lack of the term
 $\nabla \rho_h\times\nabla P_h$ prevents the thermodynamic driving of vorticity 
 by the background, which would usually lead in a baroclinic disk to phenomena like the 
 Rossby wave instability \citep{1999ApJ...513..805L}.
In shearing boxes, this kind of instability
can be driven by localized gradients, or for example, by adopting a Boussinesq approximation for the system at the point of  
Equation~(\ref{eq:mom_full_SP_compatible}) and thus retaining the background radial hydrostatic pressure gradient with
shear-periodic radial boundary conditions \citep{2010A&A...513A..60L}.

\vspace{1cm}

\bibliographystyle{yahapj}

\end{document}